\documentclass[prb,showpacs,preprintnumbers,amsmath,amssymb]{revtex4}

\usepackage{graphicx}
\usepackage{dcolumn}
\usepackage{bm}
\usepackage{subfigure}

\begin{document}
\title{Confinement of skyrmion states in non-centrosymmetric magnets near ordering temperatures}

\author{A. O. Leonov}
\thanks
{Corresponding author} 
\email{a.leonov@rug.nl}



         
\maketitle



I consider skyrmionic states in non-centrosymmetric magnets near the ordering temperatures. 
I show that the interaction between the chiral skyrmions,
being repulsive in a broad temperature range, 
changes into attraction at high temperatures. 
This leads to a remarkable \textit{confinement} effect: near the ordering temperature skyrmions 
exist only as bound states. The confined skyrmions as discernible units may arrange in different mesophases. The confinement of multiple skyrmions is a consequence of the coupling between the magnitude and the angular part of the order parameter. Thus, near the ordering transitions,  I consider the local magnetization not only as multiply twisted but also longitudinally modulated.
From numerical investigations on 2D models of isotropic chiral ferromagnets, a staggered half-skyrmion square lattice at zero and low fields and a hexagonal skyrmion lattice at larger fields are found in overlapping regions of the phase diagram near the 
transition temperature. The creation of skyrmions as stable units and their condensation into different extended textures occurs simultaneously through a rare case of an instability-type nucleation transition \cite{Felix86}.  
As well, I introduce a new fundamental parameter, \textit{confinement temperature} $T_L$ separating the peculiar region with "confined" chiral modulations from the main part of the phase diagram with regular helical and skyrmion states. 

\section{Introduction}

The peculiarities of the paramagnetic to helimagnetic transition in the intermetallic compounds MnSi and cubic polymorph FeGe have been investigated experimentally since a long time and seem to be related to the specific frustration introduced by the chiral Dzyaloshinskii-Moriya interactions. These compounds represent only two examples from the rich family of B20 magnets. The literature on the experiment in MnSi is overwhelming, this compound may be considered as a laboratory for investigation of chiral magnetic modulations near the ordering temperatures. The experiments on FeGe are still scarce.  In the following I briefly summarize the main experimental results on the chiral magnetic properties of these compounds. 

MnSi and FeGe crystallize in the B20 structure with the space group P2$_1$3 possesing no inversion symmetry and containing 3-fold axes along the <111> space diagonals and 2-fold screw axes parallel to the cube axes. MnSi \cite{Kadowaki82,Lebech95} orders magnetically at 29K and forms the helical modulation with the period of 18 nm. In another P2$_1$3-compound FeGe \cite{Lebech89} the ordered phase is observed below 280K: the propagation directions of the spirals point  to <100> crystallographic axes and change to <111> with decreasing the temperature below 220K; the period of the spiral with 70 nm is much larger than in MnSi. 

\begin{figure}
\centering
\includegraphics[width=18cm]{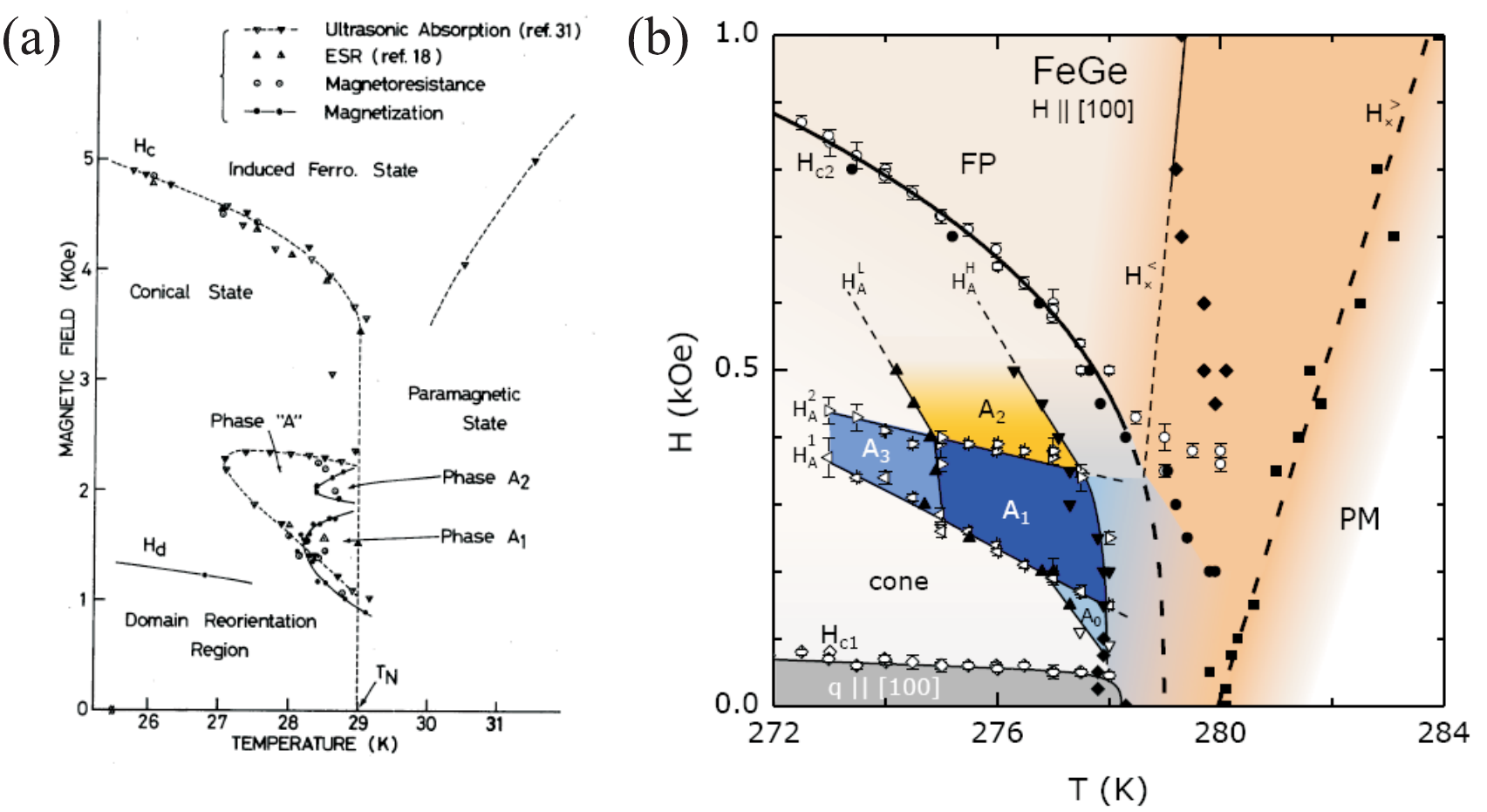}
%
\caption{
\label{PDexp}
(a) Phase diagram of MnSi near the ordering temperature found by different techniques (after Kadowaki \cite{Kadowaki82}). (b) Phase diagram of FeGe with the magnetic field $H||[100]$ (after H. Wilhelm [XIII]).
}
\end{figure}

Intensive long-term experimental investigations of the chiral helimagnet MnSi and new experiments on FeGe report numerous physical anomalies along the magnetic ordering transition, and particularly, indicate the existence of a small closed area in ($H,T$) phase diagram, the so-called "A-phase" \cite{Kusaka76,Komatsubara77,Kadowaki82,Gregory92,Thessieu97,Lamago06,
Neubauer09,Ishikawa84,Lebech95,Grigoriev06,Grigoriev06a,Muhlbauer09}. As the modulation period is of the order of several hundred unit cells, the small angle neutron scattering is an appropriate tool to reveal different precursor anomalies. The first neutron scattering experiments on MnSi have been performed by Ishikawa \cite{Ishikawa76} in 1976. In 1984 Ishikawa and Arai \cite{Ishikawa84} interpreted the A-phase in MnSi as a paramagnetic state as no magnetic satellites around the nuclear peak were observed. Later, Lebech and Harris \cite{Lebech95} revealed that  the system in the A-pocket is still in a modulated state with the propagation direction perpendicular to the field. The magnetic phase diagram of MnSi close to the ordering temperature has been studied by different techniques, and the results were summarized by Kadowaki \cite{Kadowaki82} (Fig. \ref{PDexp}). Such a phase diagram based on ultrasound attenuation \cite{Kusaka76}, ESR \cite{Date77},  magnetization and magnetoresistance \cite{Kadowaki82} suggest the subdivision of the A-phase in different phases with  phase separation lines between them. At least two subregions, A$_1$ and A$_2$, can be singled out within the A-phase. Recent neutron scattering experiments revealed the six-spot pattern within the A-pocket \cite{Muhlbauer09} due to magnetic modulations transversal to the applied field and rather independent on the field direction. However, such  "six spots" patterns as well as the Hall effect measurements \cite{Munzer10} do not give the  direct evidence for the existence of skyrmion states in the A-phase of MnSi. Note, that direct observation of skyrmions in nanolayers of FeGe \cite{Yu10a} and (Fe,Co)Si \cite{Yu10}  have been performed at temperatures far below the Curie ferromagnetic temperature and is not relevant to the A-phase.   
The phase diagram of FeGe near the ordering temperature also displays a complex structure of A-pockets with a complex succession of temperature- and field-driven crossovers and phase transitions [XIII]. 

In zero magnetic field the precursor effects have been observed above $T_C$ as a ring of diffuse scattering in small angle neutron scattering on MnSi \cite{Ishikawa84,Ishikawa85,Lamago06} and FeGe \cite{Lebech89}. In both cases the radius of the ring is equal to the ordering wave vector.  By doing neutron spin-echo spectroscopy \cite{Pappas09} in MnSi such precursor effects were attributed to a  non-trivial "spin-liquid phase" appearing in a limited temperature range just above T$_C$.

Described experiments show that the modulated phases mediating  between paramagnetic and helical state in zero field or with applied magnetic field  (the A-phase region) cannot be associated with a distinct simple phase. The A-phase must be explained by certain different  mesophases. In this sense the current theoretical attempts to explain the A-phase by the formation of a specific modulated phase either with a one-dimensional (1D) modulation ("single"-$q$ helicoids) \cite{Maleyev10,Grigoriev06,Grigoriev06a} or as "triple-$q$" modulated textures \cite{Muhlbauer09} are considered to contradict the experimental data and a theoretical analysis of this chapter [XII,XIV,XV].

Within the phenomenological theory developed in the present chapter (section \ref{PTconfinement}), the occurrence of an anomalous precursor regime in chiral magnets near magnetic ordering relies on the formation of localized (solitonic) states \cite{Nature06}[XIX,XV] (section \ref{Solutionsconfinement}). Theoretical analysis of localised 1D and 2D magnetic states in a cubic helimagnet reveals that the interactions between helicoidal kinks \cite{Schaub85}, and skyrmions [XII,XIV,XV] near the ordering temperature become attractive (section \ref{Crossover}). The strong coupling between angular degrees of freedom of magnetization to its longitudinal magnitude has the consequence that magnetic ordering occurs as the simultaneous nucleation and condensation of stable solitonic units (section \ref{Confinedskyrmions}). The temperature range below ordering, where this coupling of longitudinal and transversal magnetization components occurs, is determined by a characteristic {\em confinement temperature} $T_L$ [XII,XIV,XV] (section \ref{Phenomenonconfinement}). These phenomena have a universal character and are relevant near the paramagnetic state [XII,XIV,XV].
The details of the mesophase formation  sensitively depend on very small additional effects such as, e.g., magnetic anisotropies, dipolar interactions, fluctuations etc. These competing influences provide mechanisms for complex and unconventional magnetic phase diagrams.  Already for low temperatures as shown in chapter 4, the minor anisotropy energy contributions cause the thermodynamical stability of modulated phases with respect to each other.
Some characteristic features of the experimental phase diagrams Fig. \ref{PDexp} can be interpreted also in the framework of the modified Dzyaloshinskii model for metallic chiral cubic helimagnets \cite{Nature06} (section \ref{NonHeizenberg}) that neglects such secondary effects 
but may be able to provide a more realistic description of the inhomogeneous twisted magnetic structure in these mesophases.


\section{Phenomenological theory and equations \label{PTconfinement}}

Near the ordering temperatures the magnetization amplitude varies under the influence of the applied field and temperature. 
Commonly this process is described by supplementing the magnetic energy  with an additional term $f_0(\mathbf{M})$ \cite{Bak80}. 
%
%
By rescaling the spatial variable, the magnetic field,    and the magnetization 
\begin{eqnarray}
\mathbf{x} = \frac{\mathbf{r}}{L_D},\,\mathbf{h} = \frac{\mathbf{H}}{H_0},\,\mathbf{m} = \frac{\mathbf{M}}{M_0}
\label{units1}
\end{eqnarray}
where 
\begin{equation}
L_D = \frac{A}{D},\, H_0 = \kappa M_0, \,M_0 = \sqrt{\frac{\kappa}{a_2}},\, \kappa = \frac{D^2}{2A},
\end{equation}
energy density can be written in the following reduced form
\begin{eqnarray}
\Phi   =(\mathbf{grad}\: \mathbf{m})^2 - w_D (\mathbf{m}) -\mathbf{h} \cdot \mathbf{m} + a m^2+ m^4.
\label{HTdens}
\end{eqnarray}
Coefficient $a$ is expressed as
\begin{equation}
 a = \frac{a_1}{\kappa} = \frac{J(T-T_c)}{\kappa}.
 \label{coeffa}
 \end{equation}
Alongwith three internal variables (components of the magnetization vector $\mathbf{m}$) functional (\ref{HTdens}) includes  only two control parameters, the reduced magnetic field amplitude, $h$, and the  "effective" temperature $a(T)$ (\ref{coeffa}).
By direct minimization of Eq. (\ref{HTdens}) one can derive one-dimensional (helicoids and conical helices) and two-dimensional skyrmions (isolated and bound states) for arbitrary values of the control parameters. 
As in the chapter 4, I analyse first solutions for localized isolated skyrmions.

\section{Solutions for high-temperature isolated skyrmions \label{Solutionsconfinement}}

The structure of isolated skyrmions near the ordering temperature is characterized by the dependence of the polar angle $\theta(\rho)$ and modulus $m(\rho)$ on the radial coordinate $\rho$ (in chapter 4 only angular order parameter $\theta(\rho)$ has been considered).
The total energy  $E$ of such a skyrmion (per unit length along $z$) after substituting $\psi(\phi)$ (see dependences $\psi(\phi)$ corresponding to different crystallographic classes in chapter 4) is as follows:
\begin{equation}
E = 2\pi  \int_0^{\infty} \Phi (m, \theta) \rho  d \rho 
\end{equation}
where  energy density is
\begin{eqnarray}
\Phi = m_{\rho}^2+ m^2 \left[ \theta_{\rho}^2 +\frac{\sin^2 \theta}{\rho^2} 
- \theta_{\rho} - \frac{\sin \theta \cos \theta}{\rho} \right] 
+ a m^2+ m^4 - h m \cos \theta
\label{HTdens2}
\end{eqnarray}
with a common convention $ \partial f/ \partial x \equiv f_x$.
The Euler equations for the functional (\ref{HTdens}) 
%
%
\begin{align}
& m^2 \left[ \theta_{\rho \rho} +  \frac{\theta_{\rho}}{\rho}
+\frac{\sin \theta \cos \theta}{\rho^2} 
 + \frac{2\sin ^2 \theta}{\rho} - h \sin (\theta) \right] 
 + 2 \left( \theta_{\rho} -1 \right) m_{\rho} = 0,
\nonumber \\
& m_{\rho \rho }+ \frac{m_{\rho }}{\rho}
+m \left[ \theta_{\rho}^2 +\frac{\sin^2 \theta}{\rho^2} 
+ \theta_{\rho} + \frac{\sin \theta \cos \theta}{\rho} \right] 
+ 2a m+ \nonumber\\
&\qquad\qquad\qquad\qquad\qquad\qquad\qquad\qquad\qquad\qquad\qquad+4m^3 - h \cos (\theta) =0
\label{Euler1}
\end{align}
with boundary conditions 
\begin{equation}
\theta(0) = \pi, \theta(\infty) = 0, m(\infty) = m_0, m(0)=m_1
\label{HTboundary}
\end{equation}
describe the structure of isolated skyrmions. The magnetization of the homogeneous  phase $m_0$ is derived from equation:
\begin{equation}
2a m_0+ 4m_0^3 - h =0.
\label{m0}
\end{equation}

Eq. (\ref{Euler1}) can be solved numerically. 
But before to consider typical solutions  $\theta (\rho)$, $m (\rho)$ of Eqs. (\ref{Euler1}) I consider the asymptotic  behaviour of skyrmion solutions and some remarkable results that can be obtained by simple means.
For 1D kinks such an analysis was done in Refs. \cite{Schaub85,Yamashita87}. 

\subsection{Crossover of skyrmion-skyrmion interactions \label{Crossover}}

\begin{figure}
\centering
\includegraphics[width=18cm]{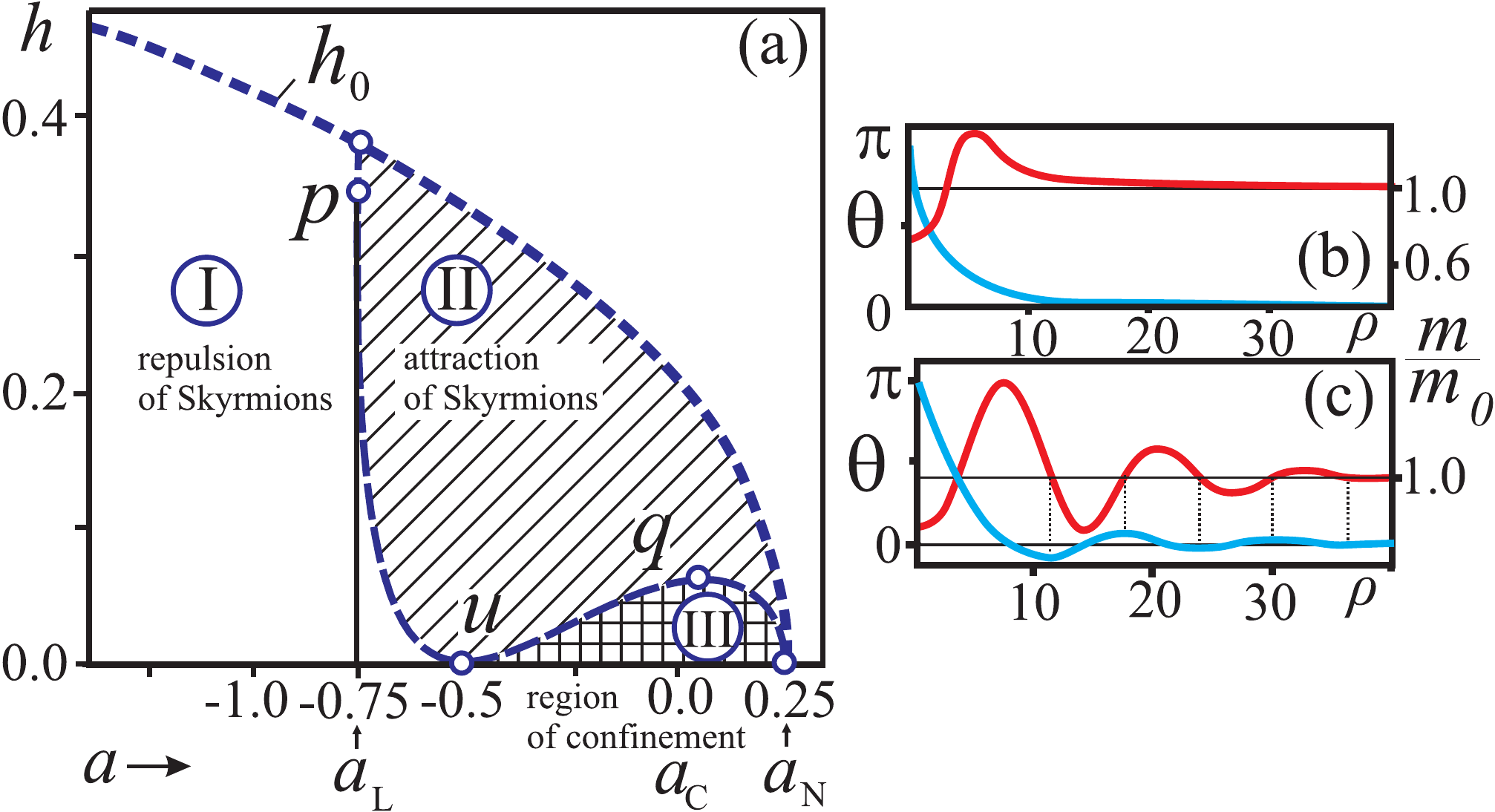}
\caption{
\label{isolated}
 The diagram on plane $(a,h)$ showing the regions with different types of skyrmion-skyrmion interaction: I - repulsive interaction between isolated skyrmions; II - attractive inter-skyrmion interaction; III - the region of skyrmion confinement. Dashed line is defined by the Eq. (\ref{criticalline1}): the turning points have the following coordinates - $p$ ($-0.75,\sqrt{2}/4$), $q$ ($0.06,0.032\sqrt{5}$), and $u$ (-0.5, 0). Above the line $h_0$ no isolated skyrmions can exist. (b) Dependences of angular $\theta$ and logitudinal $m$ order parameters on polar coordinate $\rho$  for isolated skyrmion in region I ($a=-1,\,h=0.4$). (c) $\theta(\rho)$ and $m(\rho)$  for isolated skyrmion in region II ($a=0.21,\,h=0.05$). 
}
\end{figure}

The asymptotic behaviour of isolated skyrmions bears exponential character \cite{JMMM94}:
\begin{equation}
\Delta m =(m - m_0)\propto \exp(-\alpha \rho),\, \theta \propto \exp(-\alpha \rho).
\label{assimptotic}
\end{equation}
By substituting these to the linearized  Euler equations (\ref{Euler1}) for $\rho\rightarrow \infty$
\begin{eqnarray}
\Delta m_{\rho\rho}-m_0\theta_{\rho}-\frac{1}{2}f_{mm}(m_0)\Delta m=0,\nonumber\\
m_0^2\theta_{\rho\rho}-\frac{h\theta}{2}m_0+m_0\Delta m_{\rho}=0
\label{linearized}
\end{eqnarray}
one finds three distinct regions in the magnetic phase diagram on the plane $(a,h)$ with different character of skyrmion-skyrmion interactions (Fig.\ref{isolated}):
\textit{repulsive} interactions between isolated skyrmions occur in a broad temperature range (area (I)) and is characterized by real values of parameter $\alpha\in \Re$, the magnetization in such skyrmions has always "right" rotation sense; at higher temperatures (area (II)) the skyrmion-skyrmion  interaction changes to \textit{attractive} character with $\alpha\in C$;  finally, in area (III) near the ordering temperature, $a_N=0.25$, only strictly confined skyrmions exist with $\alpha\in \Im$.

%

Equation for parameter $\alpha$ obtained from (\ref{linearized}) 
\begin{equation}
\alpha^4+\alpha^2[-2\,a-8\,m_0^2+1]+(a+6\,m_0^2)\,(a+2\,m_0^2)=0
\label{alpha}
\end{equation}
allows to write the equation for the line separating different regions: 
%
%
\begin{align}
&\left(f_{mm}(m_0)-\frac{f_{m}(m_0)}{m_0}\right)^2-4\left(f_{mm}(m_0)+\frac{f_m(m_0)}{m_0}\right)+4=0,\nonumber\\
&h=f_m(m_0).
\label{criticallinegeneral}
\end{align}
For the case of Landau expansion  the separating line looks like
\begin{eqnarray}
h^{\star} =\sqrt{2 \pm  P(a)}(a+1 \pm P(a)/2), \;
P(a) = \sqrt{3+4a}\,
\label{criticalline1}
\end{eqnarray}
with turning points $p$ ($-0.75,\sqrt{2}/4$), $q$ ($0.06,0.032\sqrt{5}$), and $u$ (-0.5, 0) (dashed line in Fig. \ref{isolated} (a)).

The typical solutions as profiles $\theta(\rho)$, $m(\rho)$ for isolated skyrmions in each region are plotted in Fig. \ref{isolated} (b)-(c).
As in the region II the exponents $\alpha$ are complex numbers, the profiles display antiphase oscillations (Fig. \ref{isolated} (c)). Rotation of the magnetization in such an isolated skyrmion contains two types of rotation sence: if rotation has "right" sense, the modulus increases, and otherwise, modulus decreases in parts of the skyrmion with "wrong" rotation sense.
Such a unique rotational behaviour of the magnetization is a consequence of the strong coupling between two order parameters of Eq. (\ref{Euler1}) - modulus $m$ and angle $\theta$.  
%

%

%
%

\subsection{Collapse of skyrmions at high fields \label{Collapse}}

\begin{figure}
\centering
\includegraphics[width=18cm]{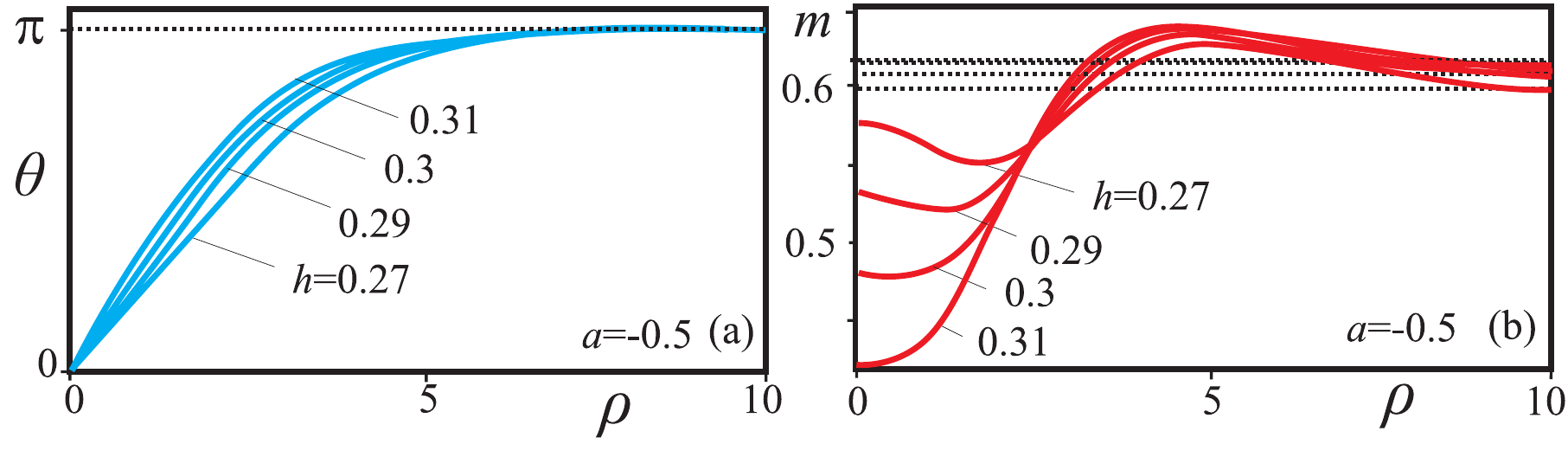}
\caption{
\label{ISMagnetization} Increasing magnetic field applied to isolated skyrmion ($a=-0.5$) localizes profiles $\theta(\rho)$ (a) and leads to the disappearence of isolated skyrmions by squeezing out modulus $m_1$ in the center (b).
}
\end{figure}

The solutions of Eqs. (\ref{Euler1}) exist only below a critical line $h_{0} (T)$ (Fig. \ref{isolated} (a)). As the applied field approaches this line, the magnetization in the skyrmion center $m_1$ (Eq. (\ref{HTboundary})) gradually shrinks (Fig. \ref{ISMagnetization} (b)), and
as $m_1$ becomes zero, the skyrmion collapses. This is in contrast to low-temperature skyrmions which exist without collapse even at very large magnetic fields \cite{JMMM94} and are protected by the stiffness of the magnetization modulus which maintains topological stability of skyrmions.
At high temperatures the softness of the magnetization amplitude allows to destroy  the core of the skyrmion by "forcing" through the magnetization vector  $m_1$  along the applied field.
The angle $\theta$ nevertheless undergoes strong localization as it was also noted for "low-temperature" skyrmions (Fig. \ref{ISMagnetization} (a)). As an example, I illustrated the magnetization process for an isolated skyrmion for $a=-0.5$ (Fig.  \ref{ISMagnetization}).

\subsection{Phenomenon of confinement \label{Phenomenonconfinement}}

The coupling of angular and longitudinal order parameters may be so strong, that oscillations in the asymptotics of isolated skyrmions do not diminish. The purely imaginary exponent $\alpha$ then reflects the region of strict confinement. For $-0.5 < a < 0.25$ line (\ref{criticalline1}) delimits a small pocket (III) in the vicinity of the ordering temperature. Within this region skyrmions can exist only as bound states and drastically differ from those in the main part of the phase diagram. 

The confinement temperature $a_L=-0.75$ subdivides the temperature interval in low- and high-temperature parts: 
%
(i) in the main part ($a < a_L=-0.75$) the rotation of the local magnetization vector determines the chiral modulation, while 
the magnetization amplitude remains constant;
(ii) at high temperatures ($a_L=-0.75 < a < a_N=0.25$) spatial variation of the magnetization modulus becomes a sizeable effect, and strong interplay between longitudinal  and angular variables is the main factor in the formation and peculiar behaviour
of chiral modulations in this region.

The \textit{confinement temperature} $a_L$ [XII,XIV,XV] provides the scale that delineates the border between these 
two regimes in the phase diagram. The characteristic temperature $a_L$ is of fundamental importance for chiral magnets. It is of the same order of magnitude as the temperature interval
\begin{equation}
(a_N- a_c) \propto \frac{D^2}{A}
\label{RationConf}
\end{equation}
  (Fig. \ref{isolated} (a)), where chiral couplings cause inhomogeneous precursor states around the magnetic order temperature (for details see Ref. \cite{Nature06}). Here, $a_c$ is the conventional Curie temperature for centrosymmetric systems. When the temperature drops below $a_c$, the energy density of a ferromagnetically spin-aligned state is the lowest one. Dzyaloshinskii-Moriya interactions with negative energy density favour the rotation of the moments. Therefore, the transition to the ferromagnetic state is preceded by a transition to modulated states at the temperature $a_N$. Due to the relativistic origin and corresponding weakness of the DM exchange the shift 
\begin{equation}
\Delta a_1 =a_N - a_c,
\end{equation}
as well as 
\begin{equation}
\Delta a_2 =a_N - a_L,
\label{DeltaA2}
\end{equation}
is small.  For MnSi $\Delta a_1$ is estimated to be 0.9 K \cite{Nature06}. The shift $\Delta a_2$ must be three times as large as $\Delta a_1$ [XII,XIV,XV].


The crossover and confinement effects arise as a generic property of the asymptotic behavior of chiral solitons at large distances from the core. These effects also apply to kinks \cite{Schaub85,Yamashita87} and Hopfions \cite{Borisov10}.

\section{The structure and properties of confined skyrmions \label{Confinedskyrmions}}

\begin{figure}
\centering
\includegraphics[width=18cm]{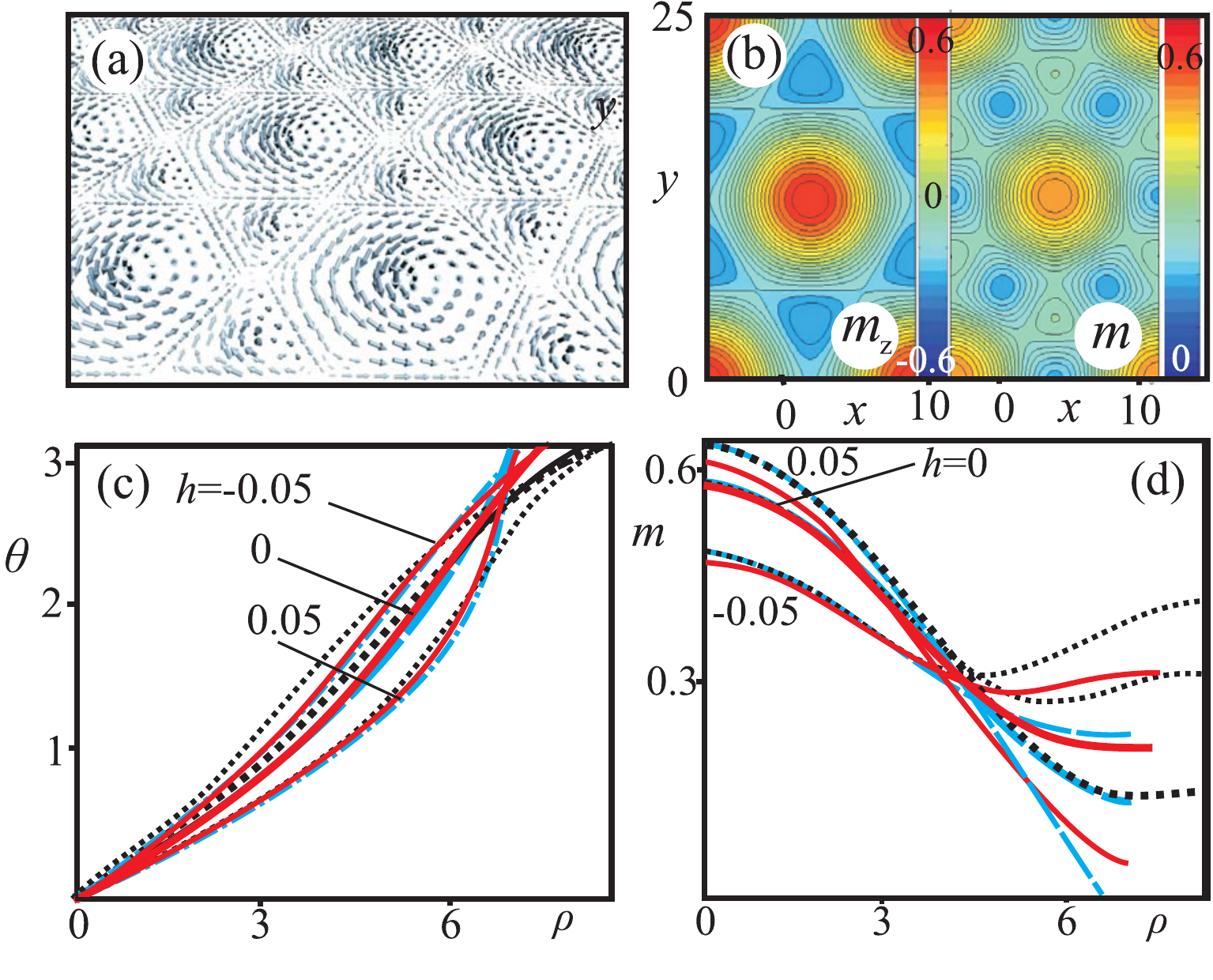}
%
\caption{The structure of the hexagonal skyrmion lattice near the ordering temperature $a_N$ ($a=0.23,\,h=0$) shown as a sketch with the distribution of the magnetization vectors (a) and the contour plots of the $m_z$-component of the magnetization and its length $m$ (b). In spite of the triangular region with the magnetization opposite to the magnetization in the "main" hexagon which are invoked to reduce the heightened energy density along the apothem of the hexagon (for details read the explanation in the section \ref{Confinedskyrmions}), the circular-cell approximation is a good approach to describe the field- and temperature-driven processes of such a lattice. (c), (d)\, Profiles $\theta(\rho)$ and $m(\rho)$ in the hexagonal cell for two different perpendicular directions through  the core (blue dashed and black dotted lines) plotted together with the solutions for skyrmion in the circular-cell approximation in the applied positive and negative magnetic field (red solid lines).
\label{example} 
}
\end{figure} 

Isolated skyrmions condense into a hexagonal lattice  below a field $h_c$ (red line in Fig. \ref{f1b}), which marks the transition between the $-\pi$-skyrmion lattice (Fig. \ref{example} (a)) and the homogeneous paramagnetic state. The typical contour plots of $m_z$-component of the magnetization and the modulus $m$ in the hexagonal skyrmion lattice near the ordering temperature are depicted in Fig. \ref{example} (b). It is seen that the central hexagon of the contour plot for $m_z$-component has been rotated in comparison with the hexagon with constant value of the magnetization length. Moreover, this hexagon is enclosed by six triangular regions with the magnetization  opposite to that in the center of the "main" hexagon. Such a redistribution of the magnetization at the outskirt of the lattice cell might be explained by the fact that the total energy density along the apothem of the hexagon   is larger than the energy along the diagonal. Therefore, the system  tries  to suppress energetically unfavourable regions by squeezing the value of the modulus [XII,XIV,XV].


The solutions for skyrmion lattices in this chapter are obtained by  using finite differences for gradient terms and adjustable grids to accommmodate modulated states with periodic boundary conditions. In addition to the angular degree of freedom, high-temperature solutions were minimized also with respect to the length of the magnetization vector in each point of the numerical grid.

As a cross-check of obtained results, I have used the circular-cell approximation. According to the method, the hexagonal cell is replaced by a circle, and one has to solve the system of differential equations (\ref{Euler1}) with the boundary conditions:
\begin{equation}
\theta(0) = \pi, \theta(R) = 0, m(R) = m_2, m(0)=m_1.
\label{boundary2}
\end{equation}
Solving these equations, where the order parameters depend only on one spatial coordinate, is essentially easier than rigorous solution for the 2D magnetization structures. In Fig. \ref{example} (c), (d) I have plotted the angular and longitudinal profiles for the skyrmions in the circular-cell approximation (red lines) and from the numerical simulations on the two-dimensional grid along perpendicular directions in the hexagon (dotted black  and dashed blue lines). The circular-cell approximation describes surprisingly well the skyrmion structures and their transformation with temperature and magnetic field  inspite the triangular regions with opposite magnetization and slight difference for the longitudinal profiles in the applied magnetic field (Fig. \ref{example} (d)).   Moreover, such an approximation allows to analyse the processes with skyrmion lattices from another perspective and to plot dependences to be hardly achieved for rigorous 2D solutions. As an example I consider the process of condensation of isolated skyrmions into the lattice.

\subsection{Condensation of isolated skyrmions into the lattice \label{CondensationLattice}}

\begin{figure}
\centering
\includegraphics[width=18cm]{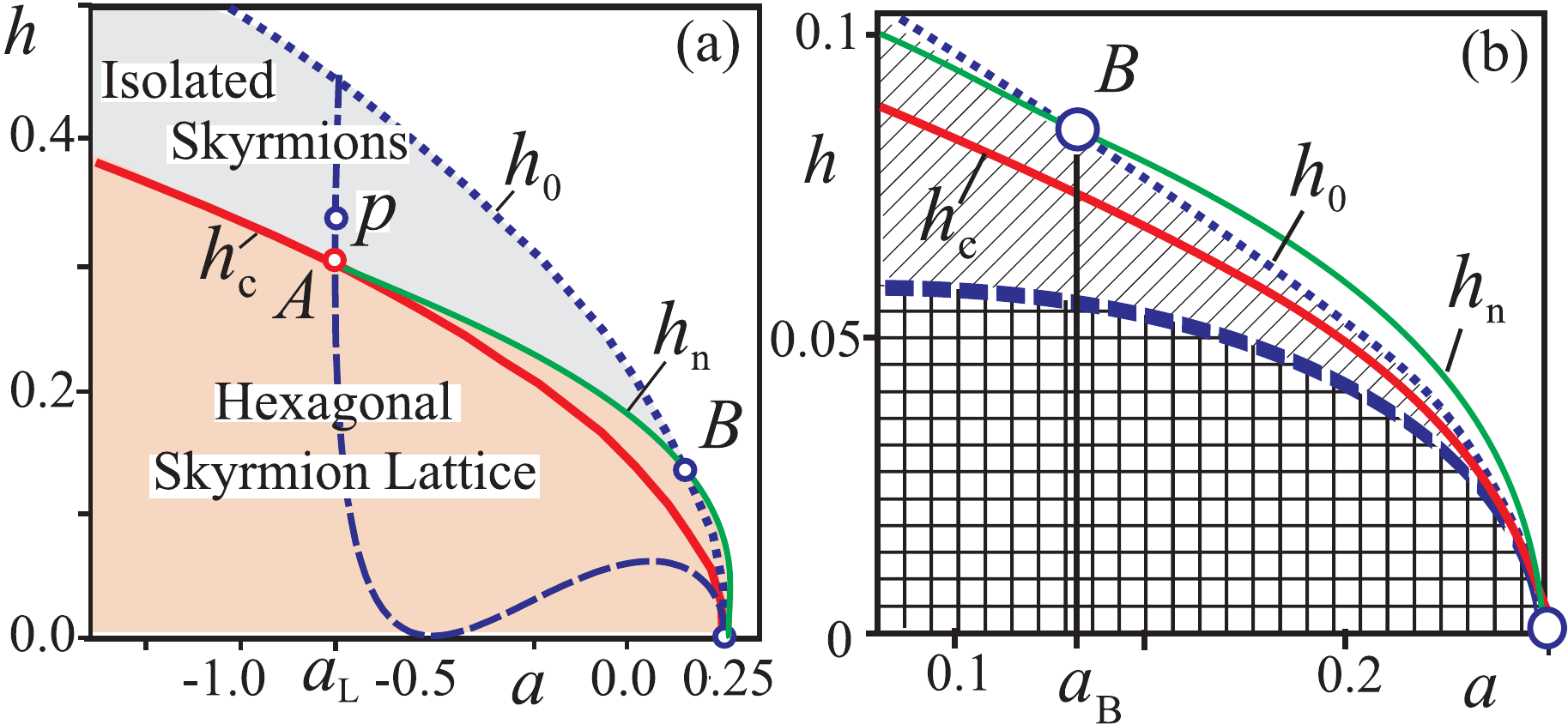}%
\caption{
\label{f1b} 
(a) Phase diagram - magnetic field $h$ vs. temperature $a$.
Below the line $h_c$ 
skyrmions condense into 
a hexagonal lattice. For $a<a_L$ the transition is second-order phase transition of nucleation type. For $a>a_L$ the transition becomes first-order phase transition. The hexagonal skyrmion lattice exists as metastable state up to the nucleation field $h_n$. 
For temperatures between points $A$ and $B$, 
$h_n < h_0$, for larger temperatures $a_B < a < a_N = 0.25$ 
the isolated skyrmions disappear at lower fields 
than the dense skyrmion lattice, $h_0 < h_n$.
For clarity, line $h_n$ is only schematically 
given in panel (a), numerically exact data 
are shown in panel (b).
}
\end{figure}

\begin{figure}
\centering
\includegraphics[width=10cm]{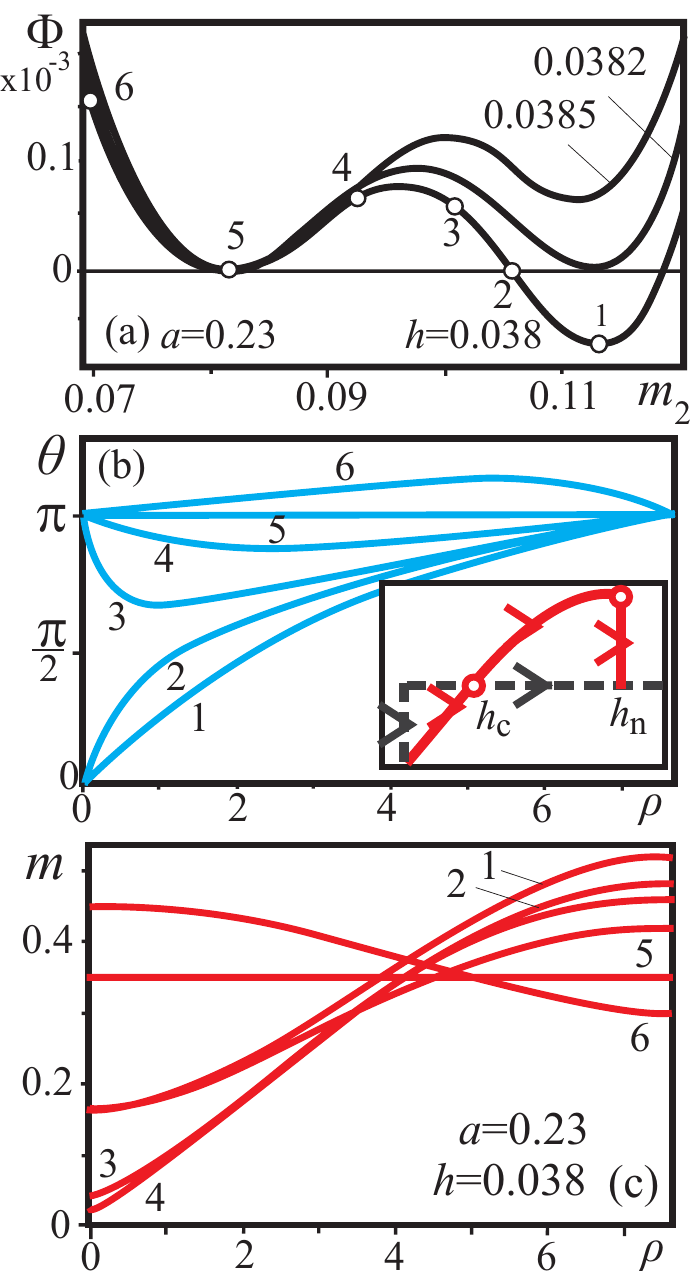}
%
\caption{
\label{hysteresis}
Energy density $\Phi$ (Eq. (\ref{HTdens2})) in dependence on the modulus $m_2$ at the boundary of skyrmion lattice cell (a) (see boundary conditions of Eq. (\ref{boundary2})) exhibits two pronounced minima corresponding to $-\pi$-skyrmion lattices (points 1) and homogeneous state (point 5). Changing the modulus $m_2$ as a parameter one can smoothly turn the hexagonal skyrmion lattice into the homogeneously magnetized state. Profiles $\theta=\theta(\rho)$ (b) and $m=m(\rho)$ (c) make it possible to trace the process of the transformation: in the intermediate points 3 and 4 the system of differential equations (\ref{Euler1}) has as a solution modulated states with the magnetization in the center and at the  periphery pointing in the same direction (in the present case, along the field); in the point 5 the homogeneous state with $\theta=0,\,m=m_0$ shows up. Such a magnetization process bears pronounced hysteretic character and takes place in the field interval $h_h<h<h_n$. 
}
\end{figure}

In the region II of the phase diagram a transition of a hexagonal skyrmion lattice into the homogeneous state is of the first order on the contrary to the second-order phase transition in the "low-temperature" region I. Thus, the hysteretic magnetization process between homogeneous state and skyrmion lattice is expected. 
The hysteretic character of the transition can be illustrated by plotting the energy density of skyrmion bound states (Fig. \ref{hysteresis} (a)) depending on the modulus $m_2$ at the outskirt of the lattice cell (see boundary conditions of Eq. (\ref{boundary2})). In the interval  of magnetic fields $h_h<h<h_n$ the energy density has two minima corresponding to homogeneous state and hexagonal skyrmion lattice, respectively. The fields $h_h$ (not shown in Fig. \ref{f1b}) and $h_n$ mark the boundaries where homogeneous state and skyrmion lattice lose their stability, respectively. On the line $h_c$ (Fig. \ref{f1b})  the energy of the hexagonal lattice equals that of the homogeneous state. In the interval $h_c < h < h_n$ the hexagonal skyrmion lattice exists as a metastable state.

Solutions for $\theta(\rho)$ and $m(\rho)$ in some points of the curve of energy density are plotted in Fig. \ref{hysteresis} (b), (c). I start from the usual skyrmion bound state (point 1) and gradually decrease the modulus on the boundary of the lattice cell. Angular profile becomes more localized (point 2) and for some value of $m_2$ there is no skyrmion lattice anymore. Instead of the hexagonal lattice the system of differential equations (\ref{Euler1}) has  a solution for a modulated state with the modulus in the center and at the outskirt pointing in the same direction (point 3 with $\theta(0)=\theta(R)=\pi$). During this process modulus in the center $m_1$ goes through zero and the magnetization in the center is reversed to the other side. Decreasing $m_2$ further, the homogeneous state is reached.

\subsection{Peculiar properties of bound skyrmions in the region of confinement \label{skyrmionProperties}}

\begin{figure}
\centering
\includegraphics[width=18cm]{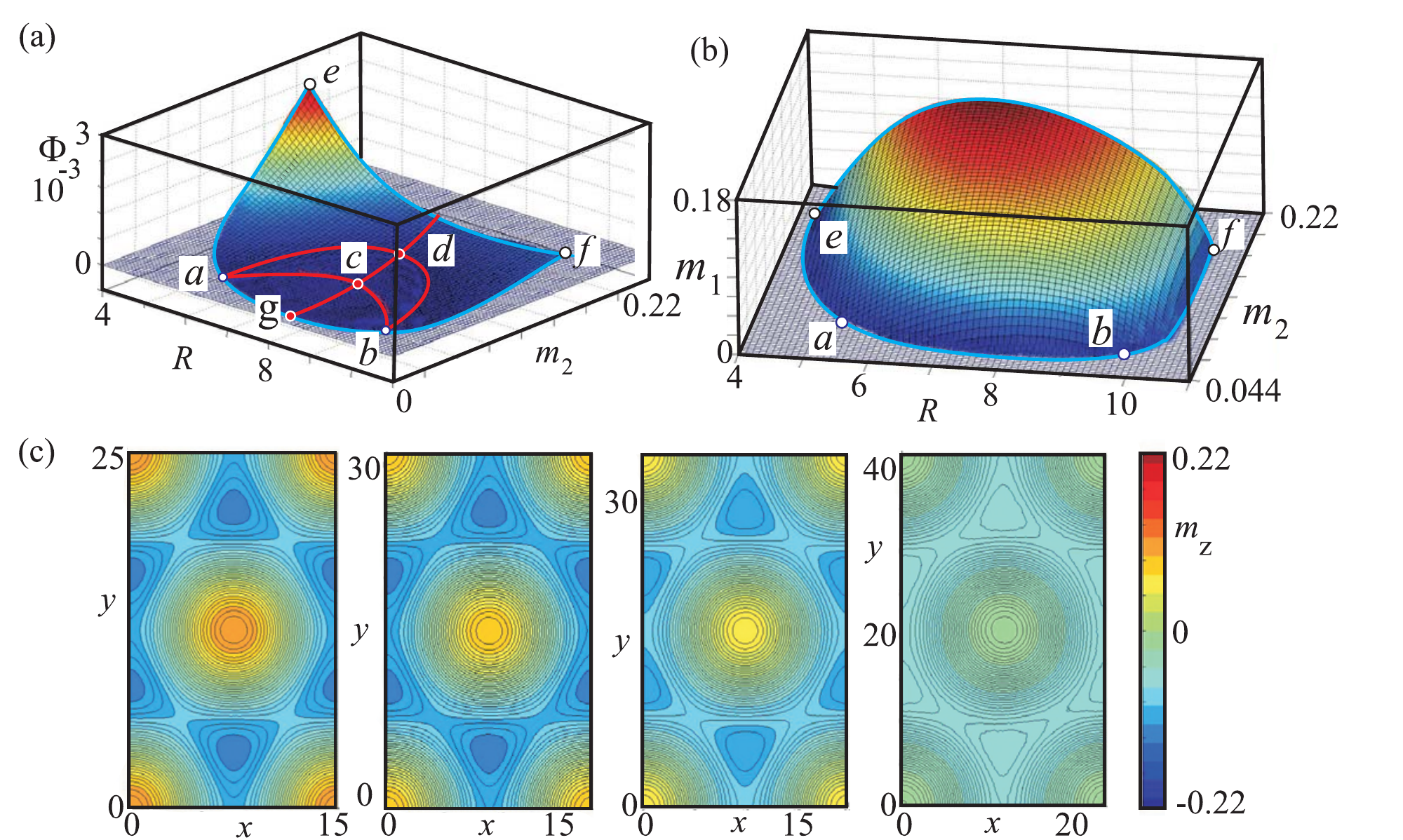}
%
\caption{
\label{expansion}
(a) Difference of energy densities of hexagonal skyrmion lattice and homogeneously magnetized state plotted as a surface in dependence on the radius $R$ of the lattice cell in circular-cell approximation and modulus at the boundary $m_2$ in correspondence with Eq. (\ref{Euler1}) and boundary conditions (\ref{boundary2}). The minimum in the point $c$ contains the equilibrium skyrmion lattice $m_z$-component of which is shown in the first snapshot of (c). (b) modulus in the center of the lattice cell plotted as a surface in dependence on the radius $R$ and  $m_2$. (c) the succession of snapshots of the skyrmion lattice  along the line $c-b$. Deviation from the minimum of energy leads to the destruction of the lattice in the point $b$ (see text for details).
}
\end{figure} 
 
As it was already noted, strong interplay between longitudinal and angular variables is the main factor in the formation and peculiar behaviour of chiral modulations in the region of confinement. The confined skyrmions  drastically differ from their "low-temperature" counterpart considered in chapter 4. Some properties of them can be described by the dependence of the radial structure on the moduli in the center $m_1$ and at the boundary $m_2$ in the circular-cell approximation as it was introduced in section \ref{CondensationLattice}. The skyrmions in the confinement region can exist only as bound states ideally as a condensed lattice. Trying to expand the skyrmion lattice, one immediately destroys it. In Fig. \ref{expansion} (a) I have plotted the  energy density of the hexagonal skyrmion lattice with respect to the homogeneous state in dependence on the radius $R$ and the modulus $m_2$ at the boundary of a lattice cell ($a=0.23,\,h=0.03$). In Fig. \ref{expansion} (b) the modulus $m_1$ in the center of skyrmion cell in dependence on the radius  $R$ and $m_2$ is depicted. 

The equilibrium state of the lattice corresponds to the minimum of the energy density (point c). For fixed modulus $m_2$ and variable radii $R$ the minimum of the energy density is located at the line $g-c-d$. In the point $d$ (and at the line $a-d-b$) the energy density becomes zero, and homogeneous state has lower energy than skyrmion lattice in the region $a-e-f-b-d-a$. In the point $g$ the modulus $m_1$ in the center of the lattice cell goes through zero, and the lattice disappears. Such a process of the lattice destruction by squeezing the modulus in the center takes place in whole at the line $e-a-g-b-f$ (Fig. \ref{expansion} (a), (b)). 

With fixed radius $R$ and variable modulus $m_2$ the minimum of energy density is localized at the line $a-c-b$. The sequence of the snapshots of the lattice exhibits the states along the line $c-b$ with the final homogeneous state reached in the point $b$.

\begin{figure}
\centering
\includegraphics[width=18cm]{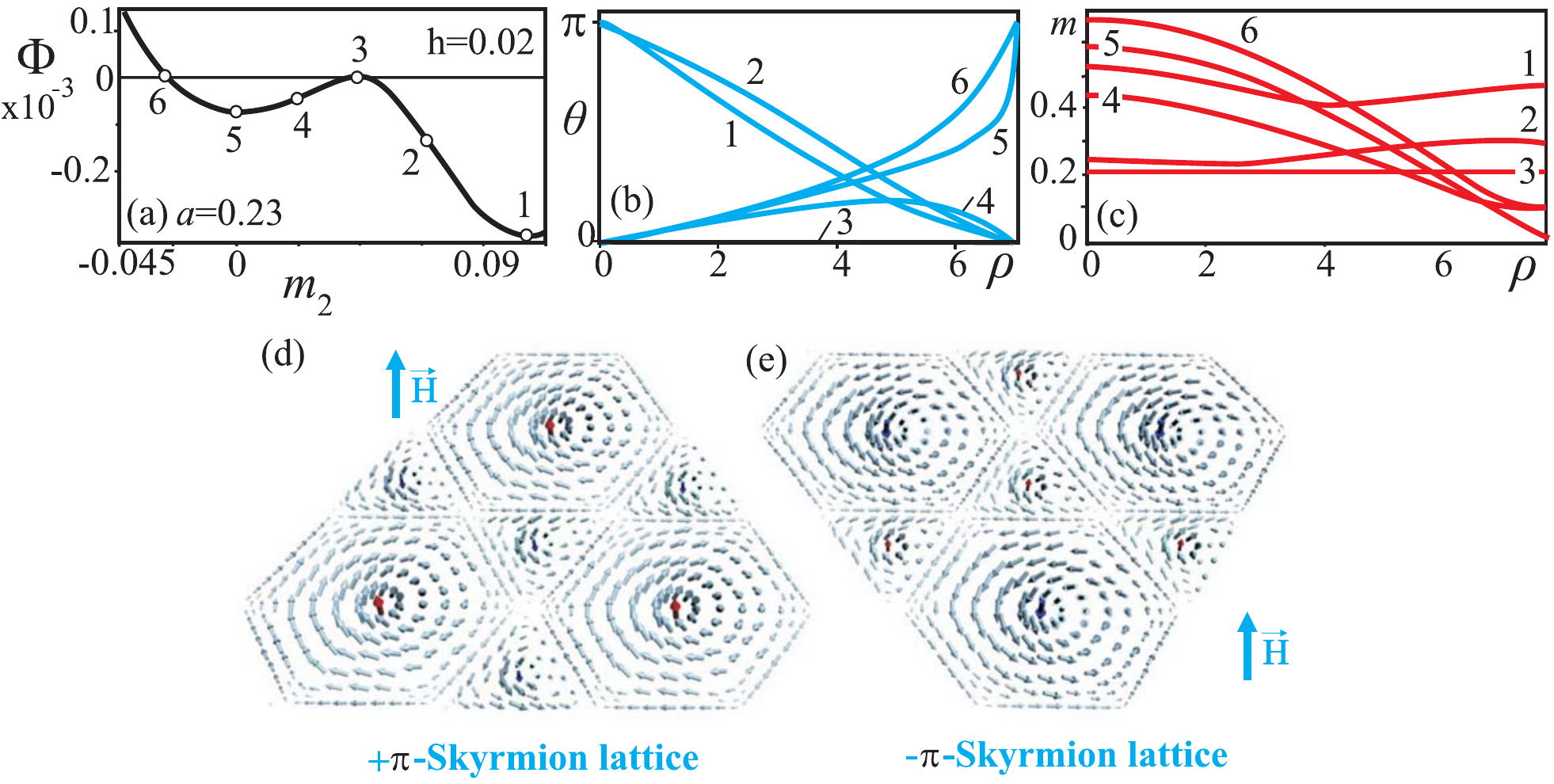}
%
\caption{
\label{MP} 
Energy density (Eq. (\ref{HTdens2})) in dependence on the modulus $m_2$ at the boundary of skyrmion lattice cell (a) exhibits two pronounced minima corresponding to $+\pi$ and $-\pi$-skyrmion lattices (points 5 and 1, respectively). Changing the modulus $m_2$ as a parameter one can smoothly turn one skyrmion lattice into the other. Profiles $\theta=\theta(\rho)$ (b) and $m=m(\rho)$ (c) make it possible to trace the process of the transformation: in the intermediate point 4 the system of differential equations has as a solution modulated states with the magnetization in the center and at the  periphery pointing in the same direction (in the present case, along the field); in the point 3 the homogeneous state with $\theta=0,\,m=m_0$ shows up. 
}
\end{figure}

By playing with the modulus $m_2$ at the outskirt  of the lattice cell  I can demonstrate also the transformation of the skyrmion lattice from the state with magnetization in the center opposite to the applied magnetic field ($-\pi$-skyrmions) to the state with the magnetization along the field ($+\pi$-skyrmions). In Fig. \ref{MP} (a) I have plotted the energy density (Eq. (\ref{HTdens2})) in dependence on $m_2$ which exhibits two pronounced minima with $\pm\pi$-skyrmion lattice in each of them. On the path from  $+\pi$-skyrmions (point 5) to $-\pi$-skyrmions (point 1) transitional structures as homogeneous state (point 3) or different modulated states with the same direction of the magnetization in the center and at the boundary (point 4) are met.

\subsection{The structure of staggered half-skyrmion lattices \label{Halfskyrmions}}

%
%

The half-skyrmion lattice depicted in Fig. \ref{half} (a) represents a special case  of modulated state near the ordering temperature which does not have "low-temperature" counterpart with a fixed modulus. A lattice of half-skyrmions is a solution for 2-dimensional energy functional (\ref{HTdens}) and has been calculated by brute-force energy minimization (it is impossible to exploit the circular-cell approximation in this case). The condensed square structure has been described as a staggered and chiral lattice
composed of half-skyrmions with four of them forming one unit conventional square lattice cell with lattice parameter $L$ (Fig. \ref{half}). These four half-skyrmions are arranged in a staggered up/down pattern. One 4th plaquette with side lengths $L/2$ of this structure is what I will call a square half-skyrmion cell. The field configuration of this half-skyrmion has the magnetization
perpendicular $m = (m_x;m_y;m_z) = (0; 0;\pm m_0)$ at the center, $r = (L/4; L/4)$. However, $|\mathbf{m}|$ is a spatially varying function with $m_0 > 0$ being the maximum value of this soft magnetization order parameter in the structure. Around this center, the  magnetization vector circulates once through the full circle. At the edge $C$ of this square cell, $m$ is in the plane. Thus, the polar angle $\theta = \arccos(\mathbf{m}\cdot \widehat{\mathbf{z}})$ of the structure rotates from 0 to $\pi/2$ from center to edge of the square cell. It is clear that this
is not the full skyrmion, it is only a piece of a skyrmion solution and it cannot exist as an isolated localized structure. The topological charge of such a cell
\begin{equation}
Q=\frac{1}{4\pi}\int_0^{L/2}\int_0^{L/2}dx dy\, \mathbf{m}\cdot[\partial_x\mathbf{m}\times\partial_y\mathbf{m}]
\end{equation}
is $Q = 1/2$. Thus, from the topological point of view, the square (sub)-cells of the lattice structure could be identified with merons \cite{Ghosh98}. The value of $Q$ depends on the integration region being only a finite part of the 2D plane. Only the definite choice of the square cell fixes this value of $Q$ unambiguously. Following terminology proposed, e.g., by Rajaraman and co-workers \cite{Ghosh98}  I call the textures in such square cells half-skyrmions. The arrangement of half-skyrmions necessitates that the magnetization $m = 0$ at $(L/2; L/2)$ passes through zero to match the half-skyrmion configurations in the interstitial
regions between the radial and chiral skyrmionic cores. Here, the corresponding field configuration of the in-plane unit vector $m = (m_x;m_y; 0)$ is an  anti-vortex.



\section[The field-driven transformation of skyrmion lattices]{The field-driven transformation of skyrmion lattices near the ordering temperature \label{skyrmionField}}

\begin{figure}
\centering
\includegraphics[width=18cm]{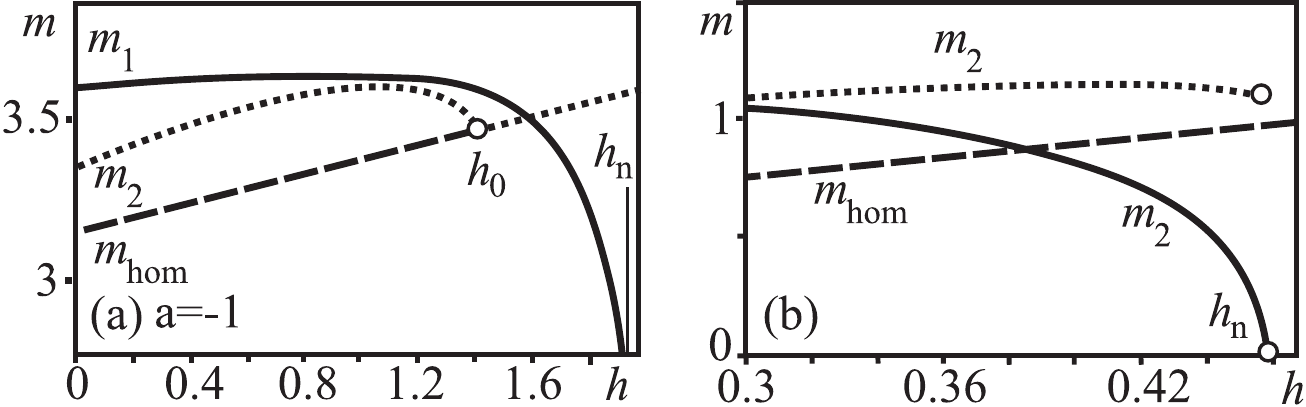}
%
\caption{
\label{moduli}
  Dependences of moduli in the center  $m_1$ and at the boundary $m_2$ of the skyrmion cell on the applied magnetic field $h$ for two values of reduced temperature $a=-1$ (a) and $a=0.14$ (b). In the first case (a) modulus $m_2$ at the outskirt "meets" the modulus of the homogeneous state $m_{\mathrm{hom}}$ for $h=h_0$. The period of the lattice in this point expands infinitely and only isolated skyrmions can exist. With $h=h_n$ marking the transition into homogeneous state, the modulus $m_1$ in the isolated skyrmion goes through zero. In the second case (b) the modulus $m_1$ in the center of the skyrmion cell shrinks to zero faster than the lattice expands into the isolated skyrmions. The bound and isolated skyrmions are two distinct branches of solutions which cannot be turned each into other (the oscillations of order parameters are very pronounced for $a>a_B$).
}
\end{figure}

\subsection{Transformation  of $-\pi$-skyrmion lattice \label{FieldMinus}}

In the applied magnetic field the hexagonal $-\pi$-skyrmion lattice disappears at the lability line $h_n$: to the left side of the point B (Fig. \ref{f1b}) the lattice releases the free isolated skyrmions (the oscillations of the order parameters are almost undetectable for $a<a_B$); to the right side of the point $B$ the skyrmion lattice transforms into homogeneous state with modulus in the center of the lattice cell going through zero. Therefore, the temperature $a_B$ can serve as another characteristic landmark together with the confinement temperature $a_L$.

In Fig. \ref{moduli} I have plotted dependences of moduli in the center $m_1$ and at the boundary $m_2$ of the lattice cell on the magnetic field for two different values of reduced temperature $a=-1$ (Fig. \ref{moduli} (a)) and $a=0.14$ (Fig. \ref{moduli} (b)) which characterize the described behavior.  

In the first case ($a=-1$) the modulus at the outskirt of skyrmion lattice becomes equal to that of the homogeneous state, the lattice expands and only isolated skyrmions can exist.  Then with increasing magnetic field, modulus $m_1$ in the center of isolated skyrmion  goes through zero and the skyrmion collapses. Such a  case corresponds to the major part of the phase diagram for $a<a_B$. The evolution of skyrmion lattices under a magnetic field closely agrees with the behavior studied earlier for the low-temperature limit \cite{JETP89,JMMM94}: the transition mechanism at the high-field limit is of the nucleation type with isolated skyrmion excitations appearing below the instability line $h_0$.

\begin{figure}
\centering
\includegraphics[width=18cm]{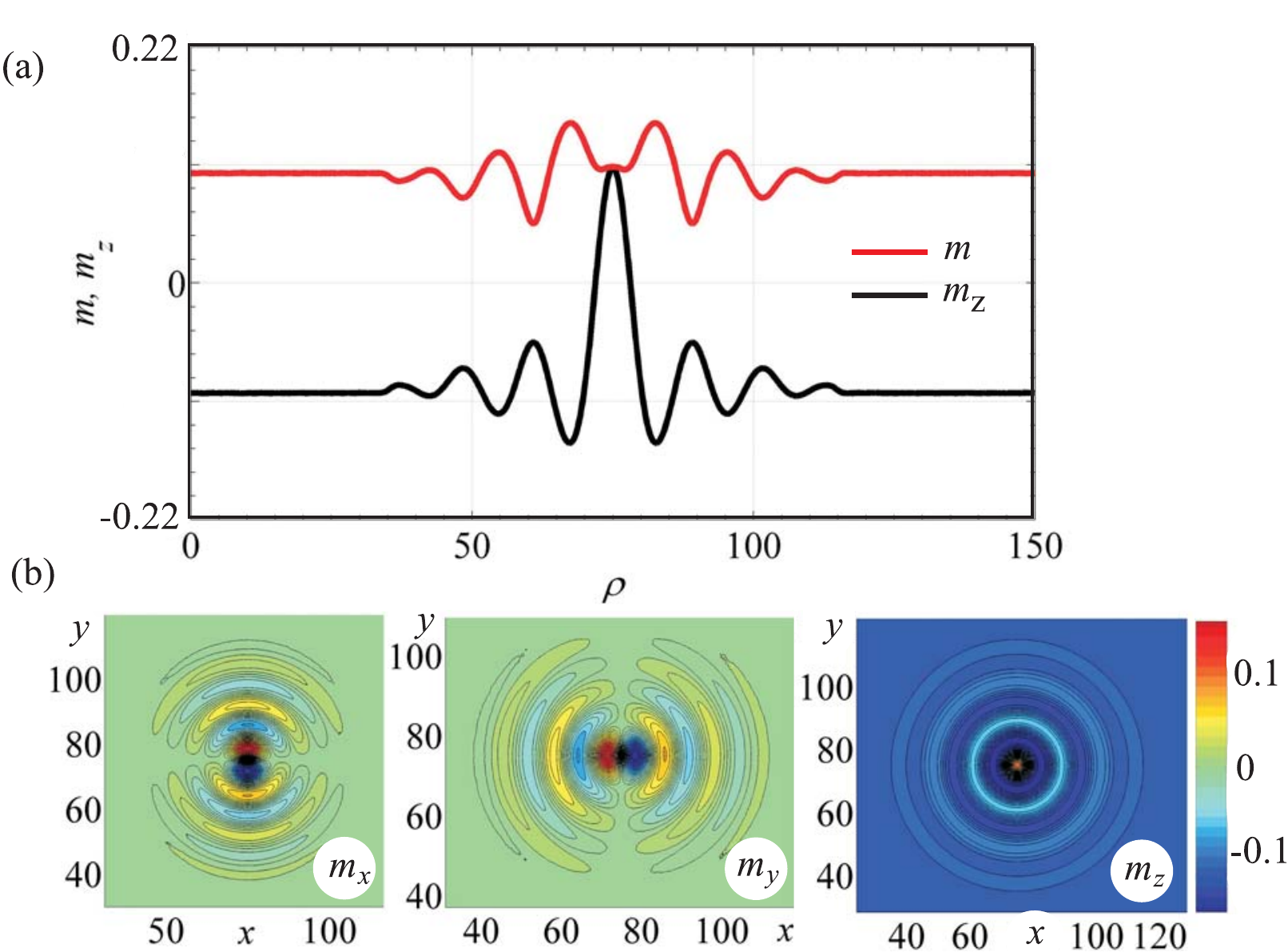}
%
\caption{
\label{OneSk} 
(a) Structure of isolated skyrmion characterized by the dependencies of the modulus (red line) and $m_z$-component of the magnetization (black line) in the cross-section for $a=0.21,\,h=-0.048$; (b) contour plots of the components of the magnetization.
}
\end{figure}

In the second case ($a=0.14$), however, the modulus $m_1$ in the center of the skyrmion goes through zero while the skyrmion lattice still is intact. Such a lattice does not set free isolated skyrmions. Isolated skyrmions exist as a different branch of solutions and cannot condense into the lattice. 
As an example of such an isolated skyrmion, I consider its structure for some parameters $a$ and $h$ where oscillations in the skyrmion asymptotics are pronounced (for instance, for $a=0.21,\,h=-0.048$). The isolated skyrmion is embedded into the  homogeneously magnetized state, and the size of the numerical grid is chosen to garantee the full decay of oscillations in the skyrmion tail. In Fig. \ref{OneSk} (a) I have plotted the dependences of modulus and $z$-component of the magnetization through the cross section of isolated skyrmion.  Contour plots of each component of the magnetization are depicted in  \ref{OneSk} (b).

\begin{figure}
\centering
\includegraphics[width=18cm]{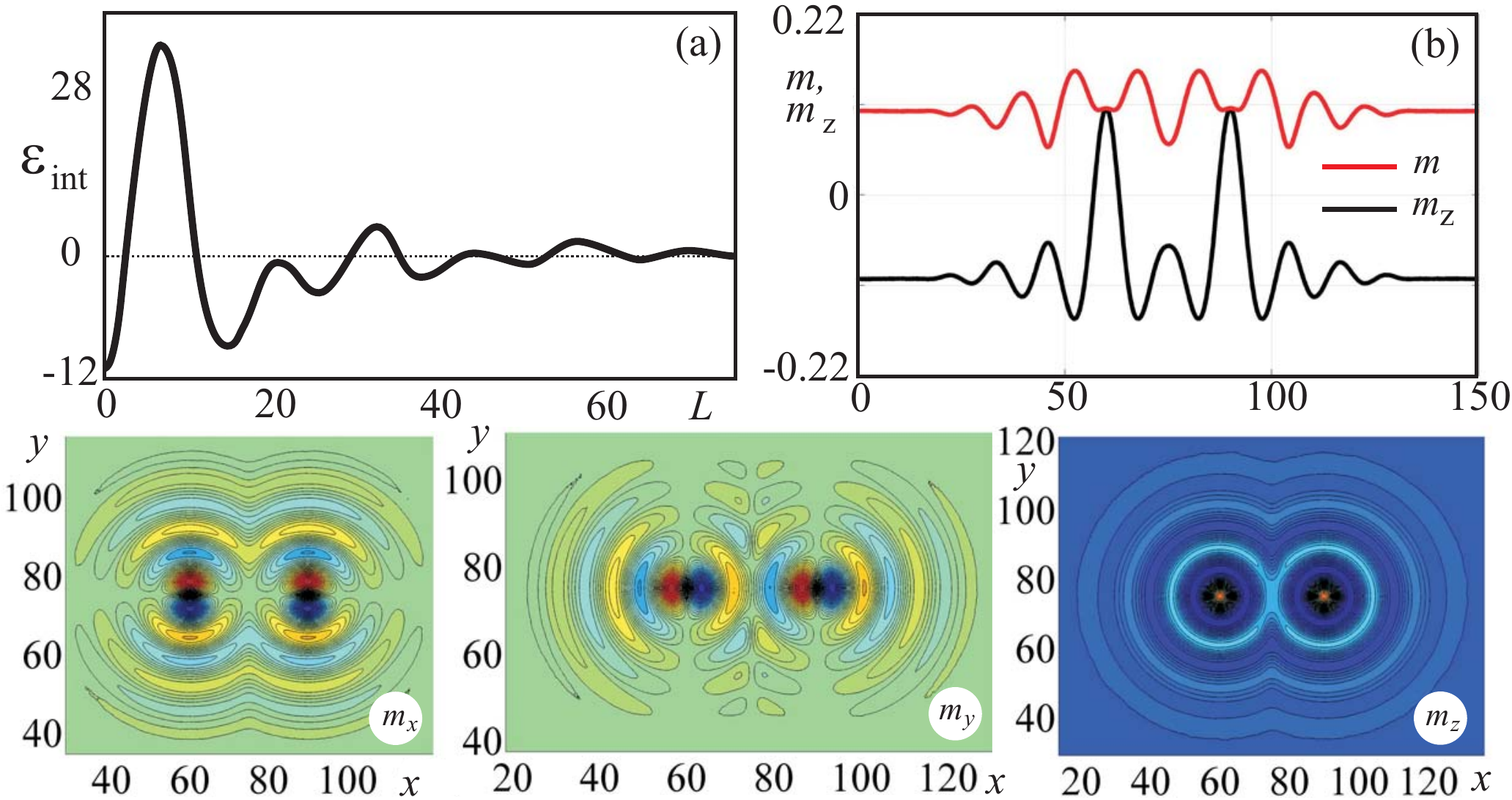}
%
\caption{
\label{TwoSk} 
(a) The skyrmion-skyrmion interaction energy $\varepsilon_{\textrm{int}}$ plotted in dependence on the distance $L$ between the centers of two isolated skyrmions. (b) Dependencies of the modulus (red line) and $m_z$-component of the magnetization (black line) in the cross-section of two interacting isolated skyrmions for $a=0.21,\,h=-0.048$ corresponding to the first (deepest) minimum of the interaction energy $\varepsilon_{\textrm{int}}$; (c) contour plots of the components of the magnetization.
}
\end{figure}

For the purpose of investigation of skyrmion-skyrmion interaction, I introduce two skyrmions into a square sample and define the interaction energy per skyrmion $\varepsilon_{\textrm{int}}$ versus the inter-skyrmion distance $L$ (Fig. \ref{TwoSk}). Due to the strongly oscillatory character of this dependence two isolated skyrmions will tend to locate at some discrete, equilibrium distances from each other and to be placed in minima of inter-skyrmion energy $\varepsilon_{\textrm{int}}$. On the other hand, single isolated skyrmion (minimum corresponding to $L=0$) cannot elongate into the pair of  skyrmions because of the high potential barrier toward the minimum with finite $L$ (the first deepest minimum of $\varepsilon_{\textrm{int}}$). Adding skyrmions one by one an optimal number of skyrmions in the cluster is found - the isolated skyrmions tend to form  the hexagonal lattice with the densest space filling. The deepest minimum of $\varepsilon_{\textrm{int}}$ for two interacting isolated skyrmions is very close to the period of hexagonal skyrmion lattice existing for the same control parameters. Importantly,  in the very narrow region between the lines $h_0$ and $h_n$ (Fig. \ref{f1b} (b)) skyrmions can exist only as bound states in the form of perfect hexagonal lattice or different cluster formations [XII,XIV,XV].

\begin{figure}
\centering
\includegraphics[width=14cm]{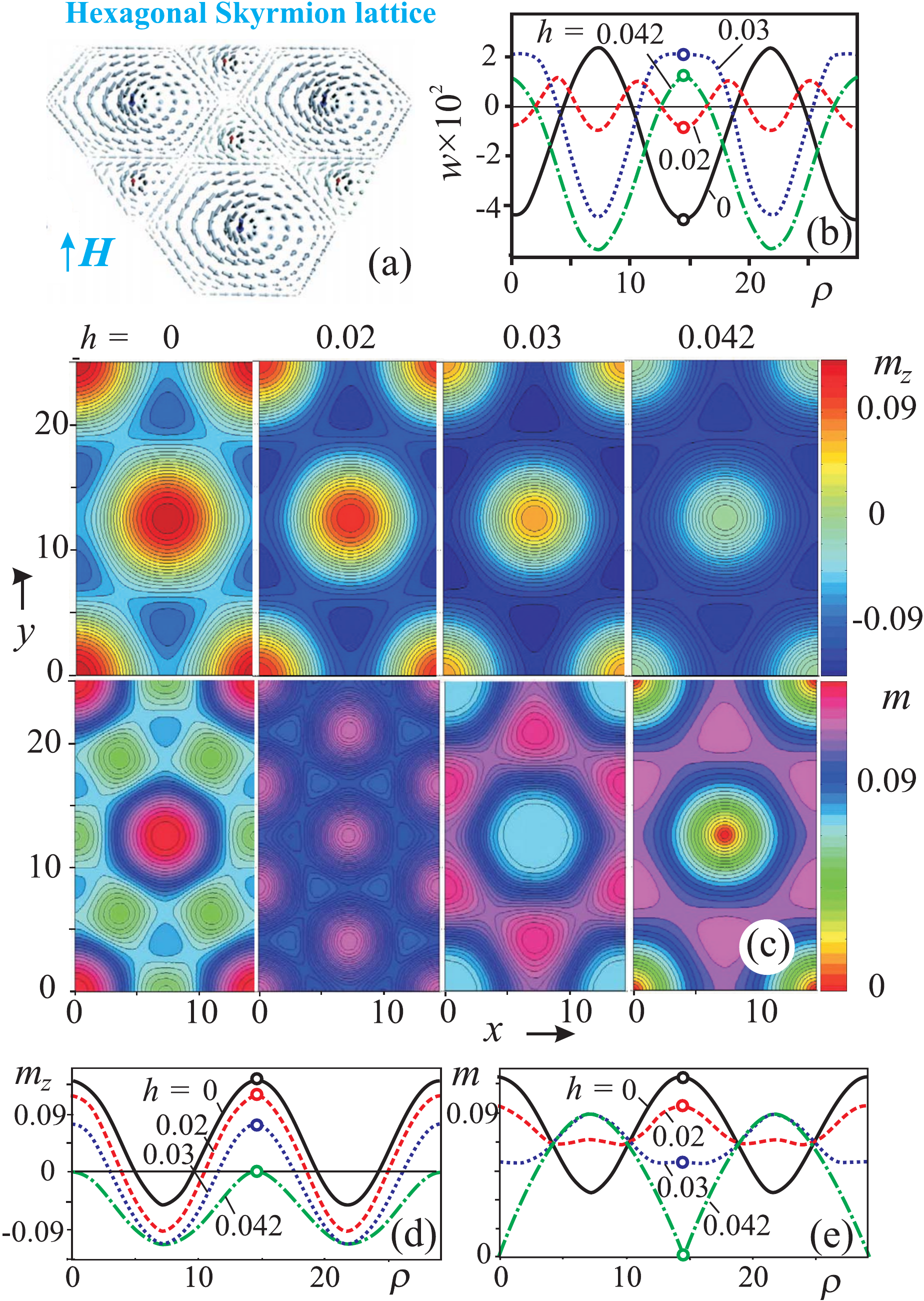}
\caption{%
\label{MinusPi}
Numerically exact solutions of  hexagonal skyrmion lattice (a) for $a=0.23$ and different values of the applied magnetic field shown as contour plots of the modulus $m$ and $m_z$-component of the magnetization (c) as well as their diagonal cross-sections (d), (e). The diagonal cross-section of the energy density is shown in (b).
} 
\end{figure} 

In Fig. \ref{MinusPi} I depict the process of transformation of the $-\pi$-skyrmion lattice into the homogeneous state for $a=0.23$ and some characteristic values of the applied magnetic field $h$.  Due to the "softness" of the magnetization modulus the field-driven transformation of the skyrmion lattice evolves by distortions of the modulus profiles $m(\rho)$ while the equilibrium
period of the lattice does not change strongly with increasing applied field. Despite the strong transformation of the internal structures the skyrmion lattice preserves \textit{axisymmetric} distribution of the magnetization near the center of the skyrmion lattice cell (Fig. \ref{MinusPi}). As for "low-temperature" skyrmions with constant modulus, the local energetic advantage  of skyrmion lattices with soft modulus over helicoids is due to a larger energy reduction in the  "double-twisted" skyrmion cell core compared to "single-twisted" helical states \cite{Nature06,Wright89}.

\begin{figure}
\centering
\includegraphics[width=18cm]{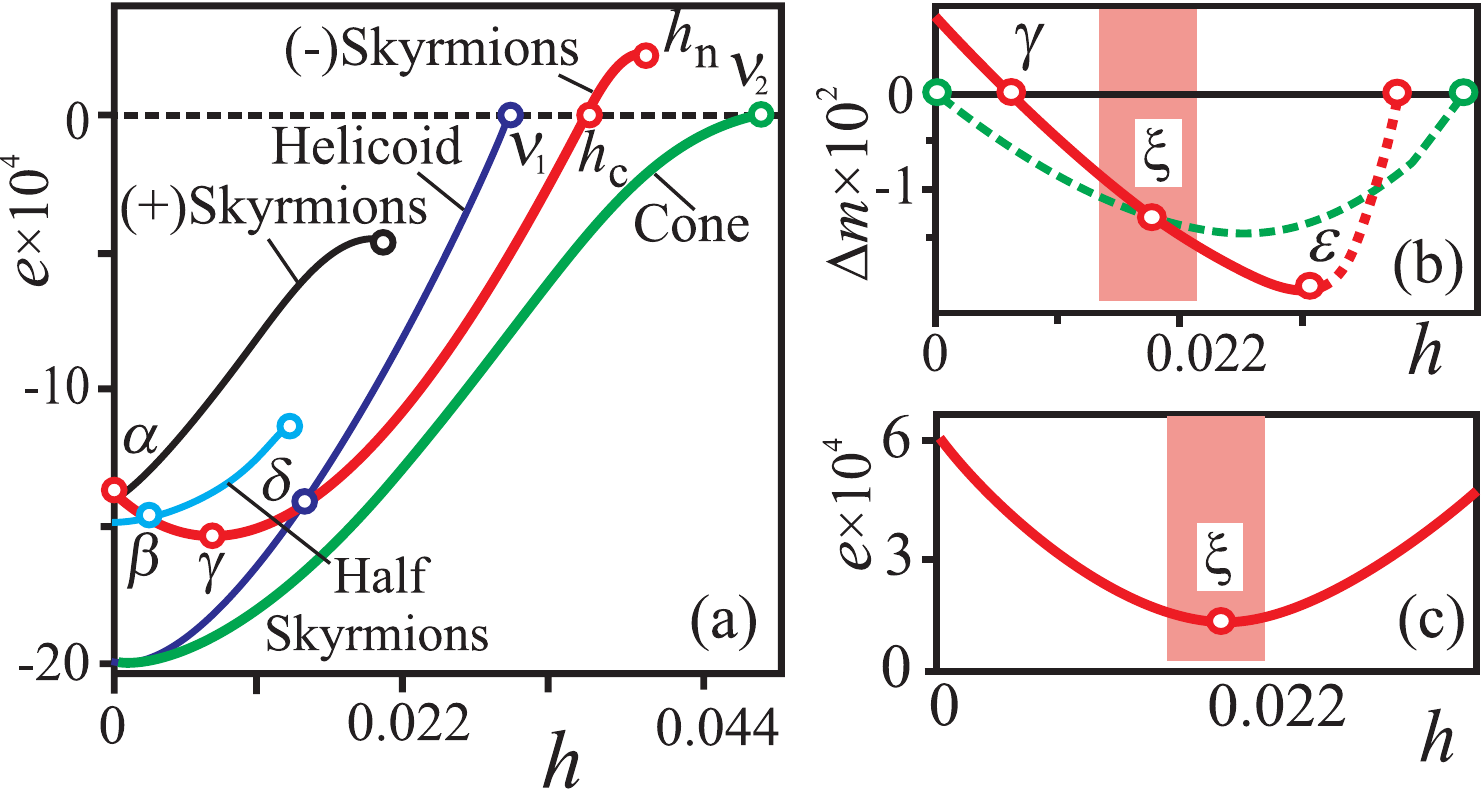}
\caption{%
\label{energies}
(a) Dependences of energy densities in all considered modulated structures on the applied magnetic field $h$ ($a=0.23$), the energy density is calculated with respect to the homogeneously magnetized state; (b) differences of the averaged magnetization along the applied magnetic field of $-\pi$-skyrmion lattice (red line) and conical phase (green dashed line) with the modulus $m_0$ of the homogeneously magnetized ferromagnetic state; (c) energy density of the $-\pi$- skyrmion lattice with respect to the conical phase exhibits the minimum  in the field (point $\xi$), where average magnetizations along the field of cones and skyrmions are equivalent.
} 
\end{figure}

The energy density of the skyrmion lattice with respect to the homogeneous state for the considered magnetization process is plotted in Fig. \ref{energies} (a). With increasing magnetic field the energy density of skyrmion lattice decreases, and in the point $\gamma$ reaches the minimum. In the minimum of the energy density the averaged value of $m_z$-components of the $-\pi$-skyrmion lattice (Fig. \ref{energies} (b)) equals  the magnetization of the homogeneous state $m_0$ (\ref{m0}).

In the point $\xi=0.02$ (Fig. \ref{energies} (b)) the average magnetization $m_z$ along the field in the skyrmion lattice intersects the magnetization of the cone. This point corresponds to the minimum of the energy difference between skyrmion lattice and conical phase (Fig. \ref{energies} (c)). The structure of the $-\pi$-skyrmion lattice in this point has a peculiar nature. The modulus in the center of skyrmion acquires the same value as in the center of triangular regions (Fig. \ref{MinusPi} (c)), and the distribution of the modulus in the skyrmion lattice becomes periodic with the doubled period as compared to initial skyrmion state in the point $\alpha$. The distribution of the energy density along $y$-direction (red dashed line in Fig. \ref{MinusPi} (b)) also indicates the special structure of the lattice in the minimum $\xi$. 

Since it is the point with the minimal energy difference with respect to conical phase, additional small energy contributions can stabilize skyrmion lattices for the field around the point $\xi$ (shaded region in Fig. \ref{energies} (c)).

The increasing magnetic field leads also to the increase of skyrmion energy density, the field gradually suppresses the antiparallel magnetization in the cell core and  reduces the energetic advantage of the "double-twist" (Fig. \ref{MinusPi} (c)). At the line $h_n$ the lability threshold of the lattice is achieved: the magnetization modulus in the cell center becomes zero (see magnetization profile  for $h = 0.042$ in Fig. \ref{MinusPi} (d), (e)). 


\subsection{Field- and temperature-driven transformation of the staggered half-skyrmion lattice \label{FieldHalf}}

\begin{figure}
\centering
\includegraphics[width=18cm]{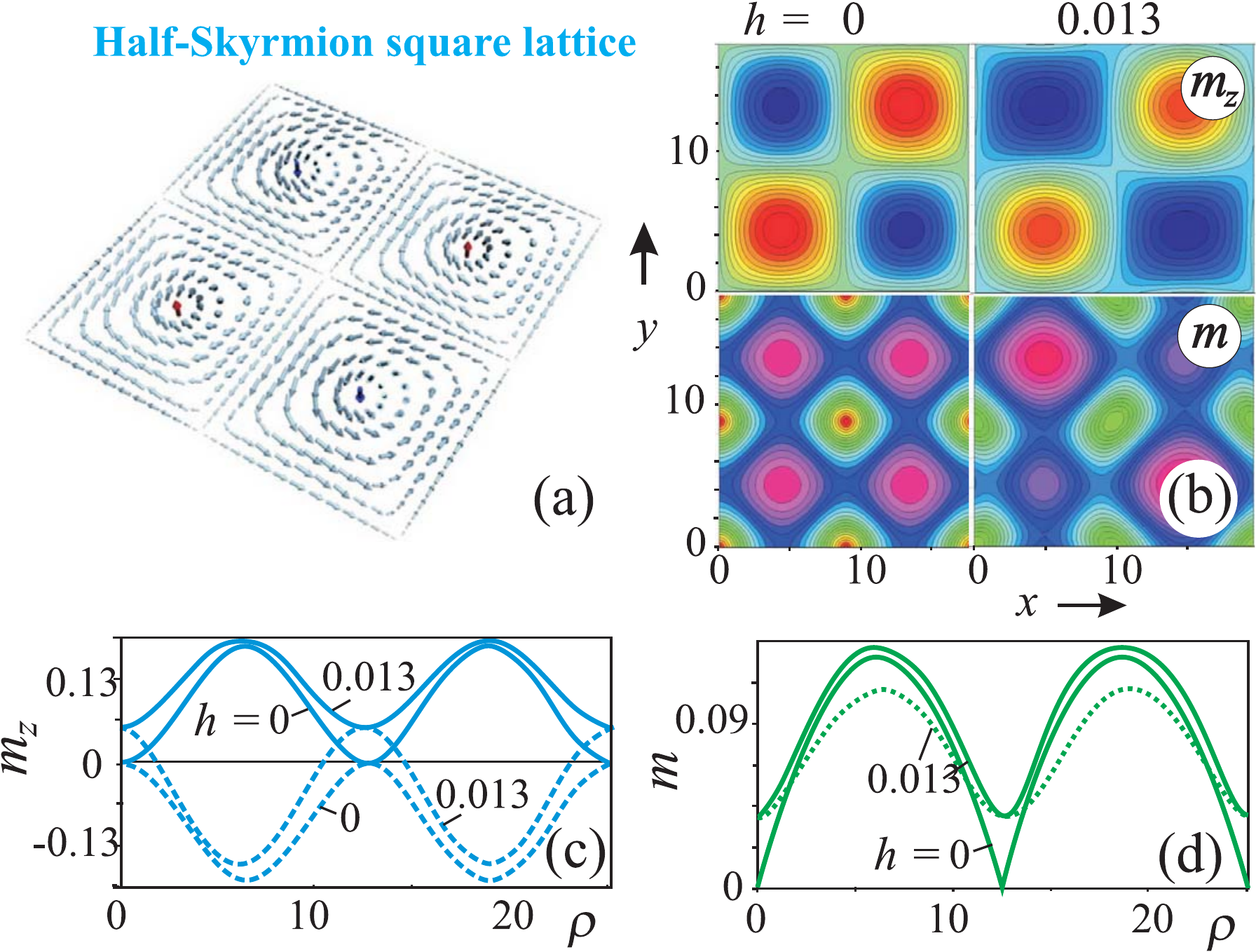}
\caption{%
\label{half}
Structure of the two-dimensional half-skyrmion lattice, derived as a minimum energy solution for the equation (\ref{HTdens}). (a) overview showing the distribution of the magnetization vectors in the base plane. (b) numerically exact solutions of  square skyrmion lattice for $a=0.23$ and different values of the applied magnetic field shown as contour plots of the modulus $m$ and $m_z$-component of the magnetization. (c),(d) the diagonal cross-sections of contour plots (b). 
} 
\end{figure} 

The dependence of the energy density on the applied magnetic field for staggered half-skyrmion lattice is plotted in Fig. \ref{energies} (a) (blue line). The energy density has a minimum relative to the homogeneous state for $h=0$ and is symmetric with respect to the direction of the applied magnetic field. 

The average $m_z$-component of the half-skyrmion lattice equals zero as it is in the homogeneous background. For $h=0$ the square half-skyrmion lattice is energetically more favourable than the densely packed hexagonal lattice, but any increase of the magnetic field leads only to the increase of its energy density. 

In the magnetic field the relative area of half-skyrmion plaquettes magnetized along the  field grows at the cost of the oppositely magnetized plaquettes.  For some value of magnetic field (point $\beta$) the energies of the square and $-\pi$ hexagonal skyrmion lattices are equivalent. Zero magnetization along the defect lines between up and down pointing plaquettes increases in the field. These interstitial regions serve as nuclei of triangular regions, since for some limiting magnetic field the half-skyrmion lattice becomes unstable and transforms into the hexagonal $-\pi$-skyrmion lattice. 
Contour plots of the $m_z$ and $m$ for $h=0.013$ exhibit already the elongation of the half-skyrmion lattice toward hexagonal one (Fig. \ref{half} (b)).

\begin{figure}
\centering
\includegraphics[width=18cm]{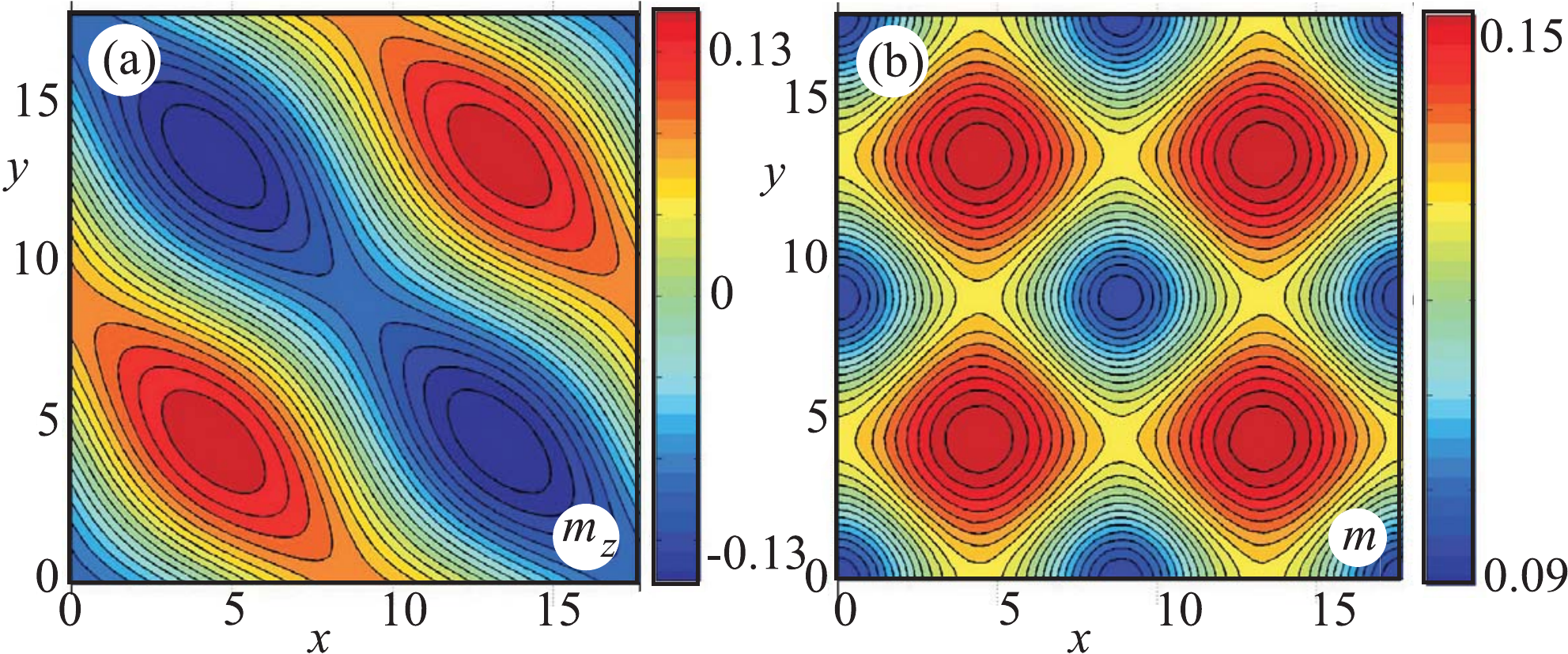}
\caption{%
\label{wave}
Contour plots of the modulus $m$ (b) and $m_z$-component of the magnetization (a) of a two-dimensional intermediate modulated structure between half-skyrmion lattice and spiral state with constant modulus. While the structure exhibits the difference of the moduli in oppositely magnetized plaquettes and maintains its square symmetry, it is apparently stable (this questions demands additional investigation). 
} 
\end{figure}

The defect lines are also the reason of the instability of half-skyrmion lattice with decreasing temperature. As it costs additional energy to make the magnetization zero along the particular line, the half-skyrmion lattice can exist only in vicinity of the ordering temperature  $a_N$. For some critical temperature $a$ the half-skyrmion plaquettes undergo elliptic instability and  elongate to form a (defected) spiral state  (Fig. \ref{wave}). The properties  of intermediate structures between the spiral and half-skyrmion lattice and the question of their stability have still to be resolved (Fig. \ref{wave}). Apparently, such a structure is stabilized by the variation of the modulus which retains the square symmetry. As the difference of the moduli in "positive" and "negative" plaquettes becomes negligible, the spiral state with constant temperature-defined modulus arises.

\subsection{Field-driven transformation of $+\pi$-skyrmion lattice \label{FieldPlus}}

\begin{figure}
\centering
\includegraphics[width=18cm]{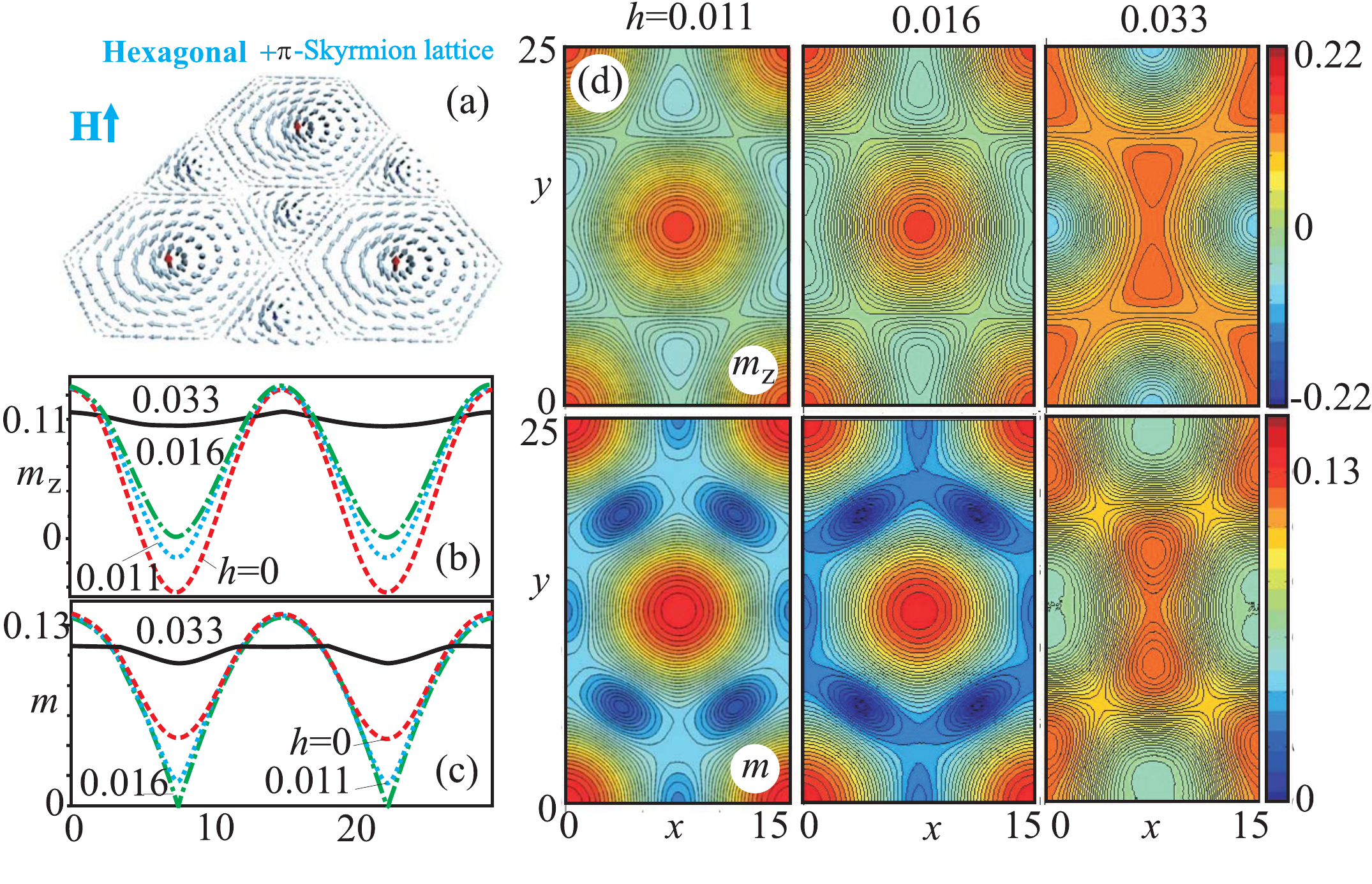}
\caption{%
\label{PlusPi} 
Structure of a two-dimensional $+\pi$-skyrmion lattice, derived as a solution for the model's equation (\ref{HTdens}). (a) overview showing the distribution of the magnetization vectors in the base plane: magnetic field is applied along the magnetization in the center of the skyrmion.  (b), (c) Distributions of the modulus and the $m_z$-component of the magnetization in the diagonal cross-sections of contour plots (d) for different values of the applied magnetic field.  (d) Numerically exact solutions of  $+\pi$-skyrmion lattice for $a=0.23$ and different values of the applied magnetic field shown as contour plots of the modulus $m$ and $m_z$-component of the magnetization characterize the process of the transformation of $+\pi$-skyrmion lattice into the more stable $-\pi$-skyrmion lattice.
} 
\end{figure} 

The $+\pi$-skyrmion lattice undergoes also a transformation toward the more stable $-\pi$-skyrmion lattice as it was described for square half-skyrmion lattice (see sect. \ref{FieldHalf}). Magnetic field applied along the magnetization in the skyrmion center stretches $m_1$ and compresses $m_2$ (see boundary conditions of Eq. (\ref{boundary2})). Such a process leads only to the increase of the energy density (Fig. \ref{energies} (a)): the $+\pi$-skyrmion lattice is the state with the largest energy density of all modulated phases under consideration.

As soon as $m_2=0$, the $+\pi$-skyrmion lattice looses its stability. In this sense, the magnetization process is reminiscent of the $-\pi$-skyrmion lattice in which $m_1=0$ in the point of the lattice instability (see sect. \ref{FieldMinus}). In increased magnetic field the modulus at the boundary of the skyrmion $m_2$ will be directed along the field. As it was explained in sect. \ref{skyrmionProperties} (Fig. \ref{MP}), such a state with the magnetization vectors pointing along the field direction in the center of skyrmion and at the outskirt represents only the intermediate state toward the energy minimum with $-\pi$-skyrmion lattice. In Fig. \ref{PlusPi} I display the process of the transformation of $+\pi$-skyrmion lattice into $-\pi$-skyrmions with the help of contour plots (Fig. \ref{PlusPi} (d)) exhibiting distribution of the modulus $m$ and $m_z$-component of the magnetization in the elementary cell as well as their dependences on the spatial coordinate in the diagonal cross-section (Fig. \ref{PlusPi} (b), (c)).


\section{Phase diagram of solutions for cubic helimagnets \label{PDcubic}}

\begin{figure}
\centering
\includegraphics[width=18cm]{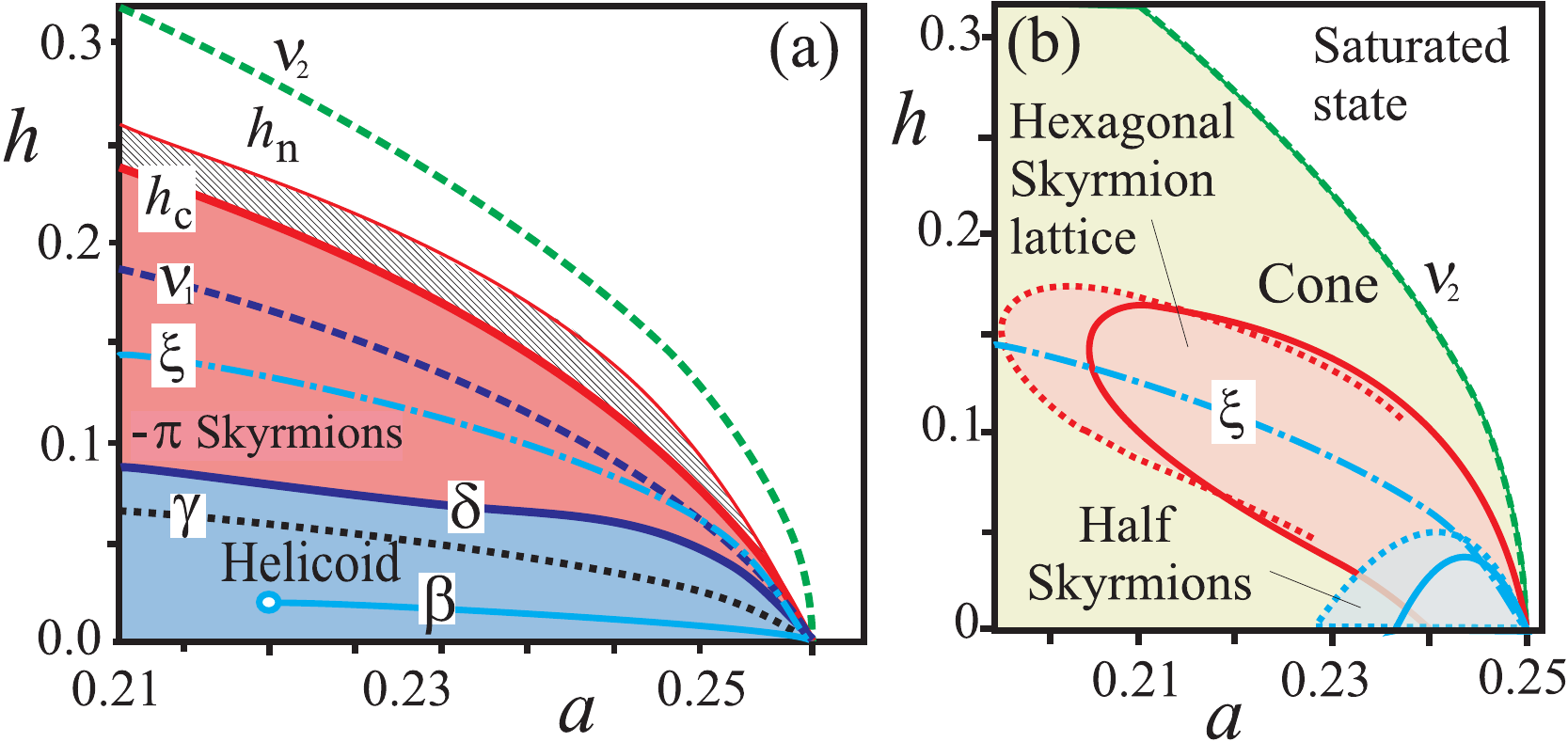}
%
\caption{
\label{stabilization}
(a) Details of the phase diagram near the ordering temperature
show the existence regions
for different modulated
states according to the energy dependences of Fig. \ref{energies} (a).
Lines for first order transitions: 
$\beta$  square half-skyrmion lattice $\leftrightarrow$ hexagonal skyrmion lattice,
$\delta$  helicoid $\leftrightarrow$ hexagonal skyrmion lattice.
Lines $\nu_1$ and $\nu_2$ mark the transition from
the helical and conical equilibrium phases into the paramagnetic phase, respectively. In the interval of magnetic fields $h_c<h<h_n$ (hatched region) the $-\pi$-skyrmion lattice exists as metastable state with respect to the homogeneous state. In the point $\gamma$ the energy density of $-\pi$-skyrmion lattice achieves minimum.
(b) Magnetic phase diagrams 
of cubic helimagnets
with exchange anisotropy
$b  = - 0.05$
and the applied field
along (111) (solid)
and (001) (dashed)
axes (a) contains
regions  with thermodynamically
stable hexagonal and square half-skyrmion lattices. 
}
\end{figure} 

In cubic helimagnets the Dzyaloshinskii-Moriya energy includes contributions with gradients along all three spatial directions. 
%
%
This stabilizes chiral modulations with propagation
along the direction of an applied field as \textit{cone} phases \cite{Bak80}.
For the isotropic model $\Phi(\mathbf{m})$
(\ref{HTdens}) the cone phase solution 
with the fixed magnetization modulus 
and rotation of $\mathbf{m}$
around the applied magnetic field:
\begin{equation}
\psi = z,\; \cos (\theta) = \frac{h}{m}, \;  m = \frac{|a-0.25|}{2},
\end{equation}
is the global energy minimum in the whole region
where modulated states exist (green line in Fig. \ref{energies} (a)).

In Fig. \ref{stabilization} (a) I plotted the phase diagram of solutions for isotropic cubic helimagnets according to the model functional (\ref{HTdens}). I showed lines for transitions between different metastable states in accordance with Fig. \ref{energies} (a). 

For cubic helimagnets, the energy density (\ref{HTdens}) 
has to be supplemented by anisotropic contributions,
\begin{equation}
\Phi_a = b \sum_{i}(\partial m_i/\partial x_i)^2 +k_c \sum_{i}m_i^4,
\end{equation}
where $b$ and $k_c$ are reduced values of exchange
and cubic anisotropies \cite{Bak80}. In the chapter 4 on the example of modulated states with the fixed length of the longitudinal order parameter $m$, I showed already  that these anisotropic interactions
impair the ideal harmonic twisting of the cone phase  and lead to the thermodynamic stability of skyrmion states. The same is true also for the modulated states with soft length of the modulus $m$ as shown in the 
equilibrium phase diagram (Fig. \ref{stabilization} (b)).

The difference 
between the energy of the hexagonal skyrmion lattice $W_{sk}$
and of the cone phase $W_{cone}$ calculated for the isotropic model,
$\Delta W_{min}= W_{sk}- W_{cone}$,
has minima along a curve $\xi(a)$. See Fig. \ref{energies} (c))
which reaches the critical point $\xi(a_N)=0$
as 
\begin{equation}
\Delta W_{min} = 0.0784 (0.25-a).
\end{equation}
Weak exchange anisotropy of a cubic helimagnet, therefore, 
creates a pocket around $a_N$, where the hexagonal 
skyrmion lattice becomes the global
energy minimum in a field (Fig. \ref{stabilization} (b)).
This case is realized in cubic helimagnets
with negative exchange anisotropy ($b < 0$)
as in MnSi \cite{Bak80}. 
This anisotropy effect provides a basic mechanism,
by which a skyrmionic texture is stabilized 
in applied fields.
Thus, the basic Bak-Jensen model \cite{Bak80} possibly can explain the observation of a skyrmion phase at finite fields in MnSi -  so-called "A-phase" \cite{Gregory92,Lebech95,Muhlbauer09}.
The exchange anisotropy $b < 0$ also leads 
to the thermodynamic stability of half-SLs (Fig.~\ref{stabilization}~(b)). 
The stabilization of these textures 
may be responsible for anomalous
precursor effects in cubic helimagnets 
in zero field \cite{Pfleiderer04,Nature06,Pappas09}.

The thermodynamic signature of the transition from the paramagnetic state into the A-phase in experiment is very similar to that into the precursor state in zero magnetic field that has been put into evidence by the observations of Pappas et al. \cite{Pappas09}. 


\section{Chiral modulations in non-Heisenberg models \label{NonHeizenberg}}

\begin{figure}
\centering
\includegraphics[width=12cm]{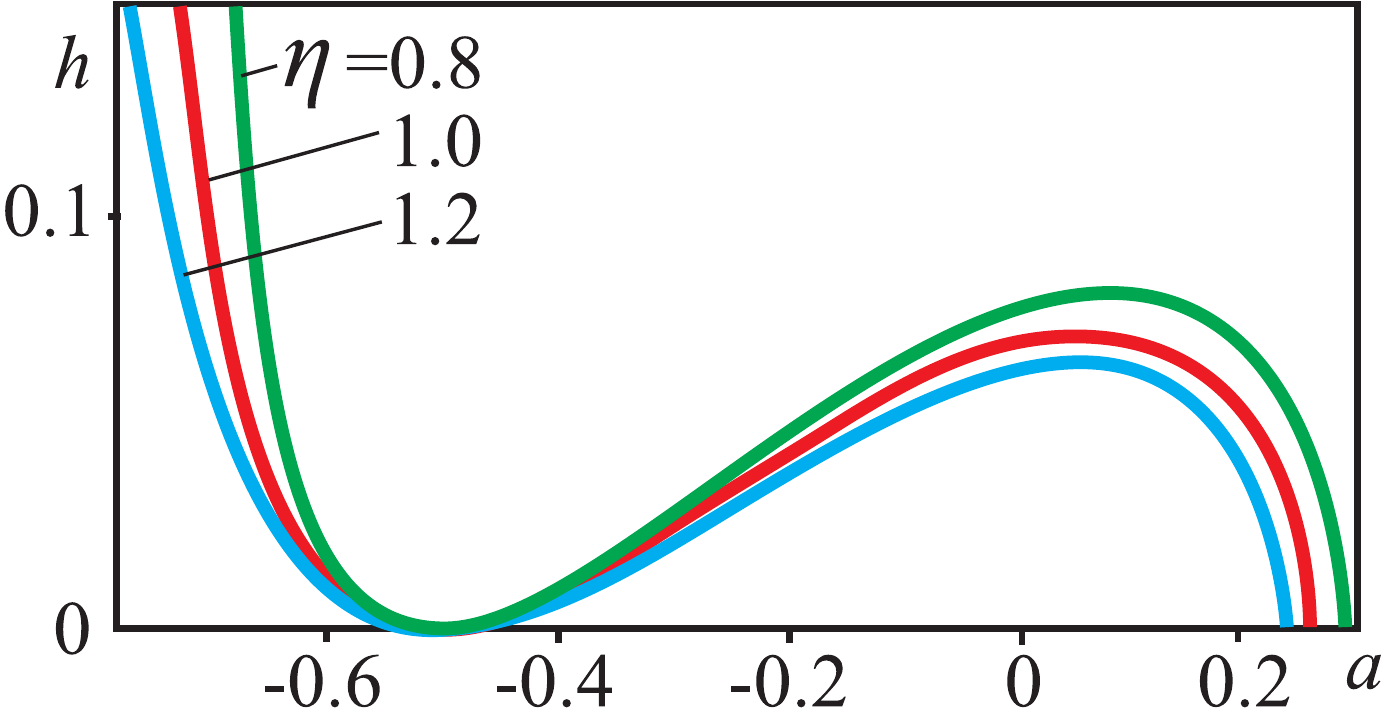}
%
\caption{
\label{eta2} Critical line (\ref{criticalline1}) in the case of non-Heisenberg model for different values of parameter $\eta$ in (\ref{etaParameter}).
 }
\end{figure}

A generalization of isotropic chiral magnets proposed in Ref. \cite{Nature06} replaces the usual Heisenberg-like exchange model 
by a non-linear sigma-model coupled to a modulus field with different stiffnesses. This yields a generalized gradient energy for a chiral isotropic system with a vector order parameter, which is equivalent to the phenomenological theory in the director formalism \cite{Wright89,Nature06} of liquid crystals:
\begin{equation}
  \sum_{i,j} (\partial_i m_j)^2 \rightarrow
 \sum_{i,j} (\partial_i m_j)^2
 + (1 -\eta) \sum_{i,j} (\partial_i m)^2
\rightarrow 
 m^2 \sum_{i,j} (\partial_i n_j)^2 + \eta \sum_i(\partial_i m)^2.
 \label{etaParameter}
\end{equation}
Parameter $\eta$ equals unity for a "Heisenberg" model, in chiral nematics $\eta = 1/3$ \cite{Wright89}. 
%

For $\eta>1$ the field- and temperature-driven evolutions of skyrmion and helical states is qualitatively the same as for $\eta=1$ (see Fig. \ref{energies} (a) for the energy dependencies of different modulated phases on the field). However for the thermodynamical stability of skyrmions,  higher values of additional anisotropic contributions must be applied. The endpoints of the lines bounding the confinement region are shifted to the left (i.e. in the region of lower temperatures) with respect to $a_N=0.25$ (blue line in Fig. \ref{eta2}). Therefore, the conical phase can exist for higher temperatures in comparison with skyrmions. 

For $\eta<1$ on the contrary, the additional "softness" of the longitudinal order parameter makes the confined chiral modulations extremely sensitive to the applied magnetic field, temperature, and anisotropic energy contributions: different chiral states undergo a very complex sequence of phase  transitions (see section \ref{magnetizationEta}).  In zero magnetic field the region of confinement extends to the temperatures higher than $a_N$ (green line in Fig. \ref{eta2}). This means that skyrmions and helicoids can exist  and compete for the thermodynamical stability for $a>a_N$. Cones appear only for $a<a_N$ independent on the value of $\eta$. 

The phase diagram of states plotted in Fig. \ref{PDeta} (a) for $\eta=0.8$ deserves a careful consideration.


%


\subsection{Field- and temperature-driven transformation of modulated states for $\eta=0.8$ \label{magnetizationEta}}

In Fig. \ref{PDeta} (b) the energy densities of all considered modulated phases are plotted with respect to the energy of the conical phase. The snapshots of the contour plots for $m_z$-components of the magnetization in particular points of these curves are shown in panels (c) and (d). These contour plots provide basic insight into the transformation of different modulated phases in the applied magnetic field. 

\vspace{3mm}
\textit{A. Transformation of the $-\pi$-skyrmion lattice in applied magnetic field }
\vspace{3mm}

For $\eta=0.8$, the hexagonal lattice of $-\pi$-skyrmions represents the metastable state with the largest energy density from all skyrmion textures.  
In the applied magnetic field the energy density of $-\pi$-skyrmion lattice increases (red line in Fig. \ref{eta2} (b), points m$_1$ and m$_2$), and eventually at some critical magnetic field $h(\mathrm{m}_3)$ the skyrmion lattice undergoes the transformation toward spiral state with the lower energy density. At the field $h(\mathrm{n}_3)$ the first-order phase transition occurs between metastable helical and $-\pi$-skyrmion states. 
To obtain numerical solution for $-\pi$-skyrmion lattice, the temperature $T_k$ of the Monte-Carlo annealing must be relatively low. Otherwise, $-\pi$-skyrmions transform into the state with the lowest energy for  $h<h(\mathrm{m}_3)$ and even for $h=0$. For $h<h(\mathrm{n}_3)$, $-\pi$-skyrmions turn into the half-skyrmion square lattice; for $h(\mathrm{n}_3)<h<h(\mathrm{m}_3)$ - into the helicoid. 

In Fig. \ref{PDeta} (d) the structure of skyrmion lattice is characterized by the contour plots for $m_z$-component of the magnetization. Magnetic field applied along the magnetization in the centers of triangular regions (blue triangles surrounding the main hexagon in Fig. \ref{PDeta} (d), points m$_1$ and m$_2$) increases significantly their fraction with respect to the parts of the lattice with opposite directions of the magnetization. In the point m$_3$ the lattice looses its stability and elongates into the spiral. In Fig. \ref{PDeta} (d) (point m$_3$) the initial moment of the transformation is shown.  

\vspace{3mm}
\textit{B. Transformation of the $+\pi$-skyrmion lattice in applied magnetic field }
\vspace{3mm}

$+\pi$-skyrmion lattice is the metastable state in the interval of magnetic fields $0<h<h(\mathrm{n}_1)$. In the point n$_1$ the first-order phase transition occurs between half- and $+\pi$-skyrmion lattices. In the interval of fields $h(\mathrm{n}_1)<h<h(\mathrm{n}_2)$, $+\pi$-skyrmions are the global minimum of the system. In the point n$_2$ the helicoids (see paragraph $D$ of the present section) replace the skyrmions  by the first-order phase transition. In the phase diagram (Fig. \ref{PDeta} (a)) the region of thermodynamical stability of $+\pi$-skyrmions is displayed by the hatching. For $h<h(\mathrm{n}_1)$ $+\pi$-skyrmions can be easily transformed into the square lattice of half-skyrmions as shown by the dotted line in Fig. \ref{eta2} (b). Therefore, the temperature of the Monte-Carlo annealing must be sufficiently low.  

In the applied magnetic field the fraction of the skyrmion lattice with the magnetization along the field grows rapidly  at the expense of the triangular regions with the opposite magnetization (point p$_1$ in Fig. \ref{PDeta} (d)). For the fields $h>h(\mathrm{n}_2)$, there are two scenarios for the evolution of this skyrmion lattice: in the first variant, the $+\pi$- skyrmion lattice turns into the helicoid as it was described also for $-\pi$-skyrmions; alternatively, $+\pi$-skyrmions may transform into the homogeneous state. 

\vspace{3mm}
\textit{C. Transformations of the half-skyrmion lattice}
\vspace{3mm}

For $\eta<1$, half-skyrmion lattice is the global minimum of the system in the interval of magnetic fields $0<h<h(\mathrm{n}_1)$ (blue line in Fig. \ref{PDeta} (b)). Additional energy costs to make  the magnetization zero along particular directions in the square lattice are lower than for $\eta>1$. As a result, the region of square lattice lability broadens essentially. For $\eta=0.8$ half-skyrmion lattice is thermodynamically stable in the temperature interval $0.152<a<0.265$, $h=0$ (see phase diagram in Fig. \ref{PDeta} (a)).

In the applied magnetic field, as it was also described in section \ref{FieldHalf}, the relative area of plaquettes in the half-skyrmion lattice magnetized along the field grows at the cost of the oppositely magnetized plaquettes (h$_2$ in Fig. \ref{eta2} (c)). For $h>h(\mathrm{n}_1)$ the half-skyrmion lattice may either transform into the more stable $+\pi$-skyrmion lattice (point h$_3$ in Fig. \ref{PDeta} (c)) with the subsequent transformation into the helicoid or elongate into the spiral state through intermediate structures shown in Fig. \ref{PDeta} (c), h$_4$. Energy density has a local minimum for such modulated states (see also Fig. \ref{wave}). 

The region of thermodynamical stability of half-skyrmion lattice is marked by blue color in Fig. \ref{eta2} (a).

\vspace{3mm}
\textit{D. Transformation of helicoids in the applied magnetic field}
\vspace{3mm}

\begin{figure}
\centering
\includegraphics[width=18cm]{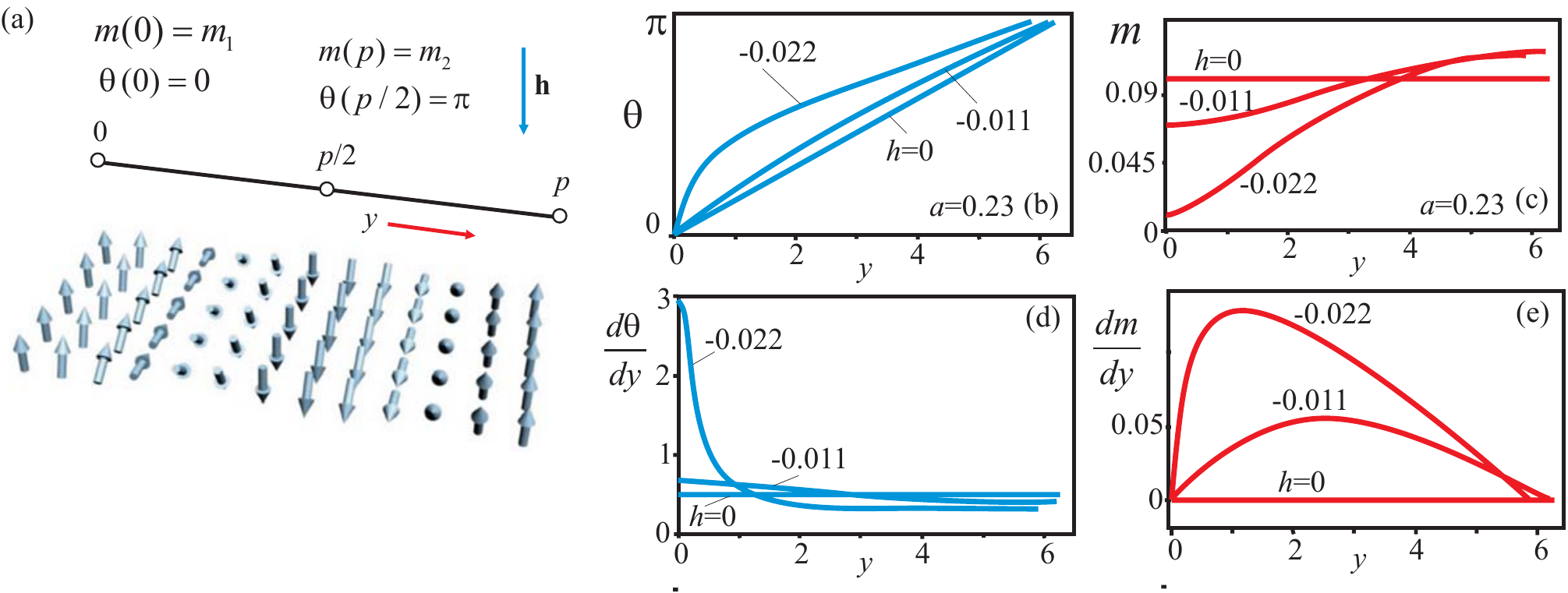}
%
\caption{
\label{SpiralEta1} Solutions for the helicoid  presented as dependences $\theta(y)$ (b), $d\theta/dy(y)$ (d), $m(y)$ (c), $dm/dy(y)$ (e) demonstrate strong transformation of the helical structure in the applied magnetic field for $a=0.23$. Longitudinal value of the magnetization along the field gradually increases, whereas opposite to the field - decreases (see sketch in (a) and longitudinal profiles in (c)). Angular profiles become more localized (see solutions in (b)). In a critical magnetic field $h=-0.024$ the magnetization opposite to the field is equal to zero, $m_1(0)=0$. 
 }
\end{figure}

For definiteness, one-dimensional helical states will be considered  to propagate  along $y$-coordinate axis; applied magnetic field is directed along $z$ (Fig. \ref{SpiralEta1} (a)). Rotating magnetization $\mathbf{m}$ is written in spherical coordinates,
\begin{equation}
\mathbf{m}=m(y)\,(\sin\theta(y),\cos\theta(y),0), \label{MagnetizationEta}
\end{equation}
with $\theta(y)$ being the angle of the magnetization  with respect to $z$ axis and $m(y)$ - the longitudinal order parameter. 

Energy density of such a helical state after substituting (\ref{MagnetizationEta}) into Eq. (\ref{HTdens}) can be written as
\begin{equation}
\Phi=m^2\left(\frac{d\theta}{dy}\right)^2-m^2\frac{d\theta}{dy}+\eta\left(\frac{dm}{dy}\right)^2+am^2+m^4-hm\cos\theta
\end{equation}

The Euler equations 
\begin{align}
&\frac{d^2\theta}{dy^2}+\frac{2}{m}\frac{dm}{dy}\frac{d\theta}{dy}-\frac{1}{m}\frac{dm}{dy}-\frac{h}{2m}\sin\theta=0,\nonumber\\
&\frac{d^2m}{dy^2}-\frac{m}{\eta}\left(\left(\frac{d\theta}{dy}\right)^2-\frac{d\theta}{dy}+a+2m^2\right)+\frac{h}{2\eta}\cos\theta=0
\label{BoundaryEta}
\end{align}
with boundary conditions 
\begin{equation}
\theta(0)=0, \,\theta(p/2)=\pi,\,m(0)=m_1,\,m(p/2)=m_2
\end{equation}
describe the structure of the helicoid in dependence on the values of the applied magnetic field $h$. $p$ is a period of the helicoid.

In Fig. \ref{SpiralEta1} (b)-(e) I have plotted the dependences $m=m(y)$ (c) and $\theta=\theta(y)$ (b) as well as $\mathrm{d}m/\mathrm{d}y(y)$ (e) and $\mathrm{d}m/\mathrm{d}y(y)$ (d) in the helicoid  for different values of the field. 
In zero magnetic field the magnetization with the constant modulus performs the single-mode rotation around the propagation direction. The longitudinal and angular order parameters are analytically defined as
\begin{equation}
m=\sqrt{\frac{0.25-a}{2}},\,\theta=\frac{y}{2}.
\end{equation}

Increasing magnetic field $\mathbf{h}||z$ destroys the single-mode character of rotation in the helicoid: magnetic field stretches the value of the magnetization along the field (m$_2$ in Fig. \ref{SpiralEta1} (c)) and squeezes it for the opposite direction (m$_1$ in Fig. \ref{SpiralEta1} (c)). The angular profiles become strongly localized (blue lines in Fig. \ref{SpiralEta1} (b)). Dependences of derivatives for corresponding order parameters are also highly non-linear (Fig. \ref{SpiralEta1} (d), (e)): the magnetization vector tries to rotate faster in the parts of the helicoid opposite to the field.

\begin{figure}
\centering
\includegraphics[width=18cm]{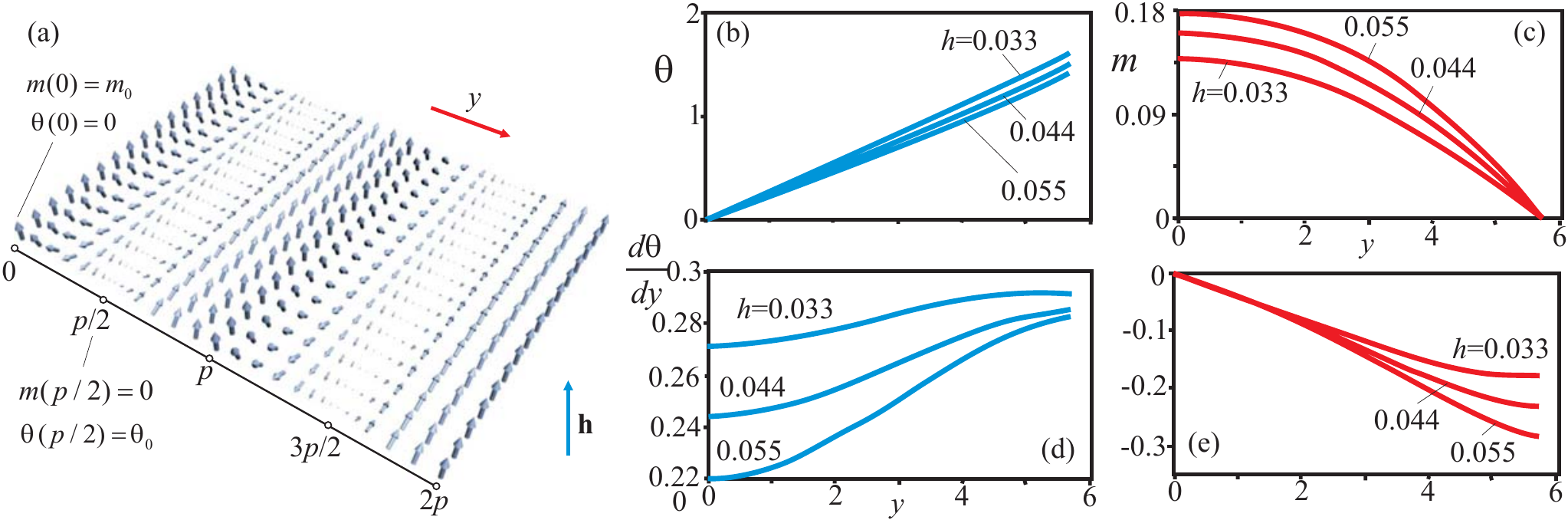}
%
\caption{
\label{SpiralEta2} Solutions for the one-dimensional modulated state with boundary conditions (\ref{PossibleBoundary}) presented as dependences $\theta(y)$ (b), $d\theta/dy(y)$ (d), $m(y)$ (c), $dm/dy(y)$ (e). Such a state is considered as a possible scenario for the evolution of a helicoid in a strong magnetic field. In (a) the structure of the helical state is presented schematically.  
 }
\end{figure}

For some critical value of the magnetic field (in Fig. \ref{SpiralEta1} for $a=0.23$, this critical field is 0.024) the value of $m_1(0)$ decreases to zero. In the further increasing magnetic field as a possible solution of Eq. (\ref{BoundaryEta}) and, therefore,  a candidate of the helicoid evolution, I considered the  one-dimensional spiral state with the following boundary conditions:
\begin{equation}
\theta(0)=0, \,\theta(p/2)=\theta_0,\,m(0)=m_1,\,m(p/2)=0.
\label{PossibleBoundary}
\end{equation}
In Fig. \ref{SpiralEta2} the same characteristic features for this spiral state as in Fig. \ref{SpiralEta1} are depicted. 
The length of the magnetization along the field continuously increases, whereas the angle $\theta(p/2)$ decreases.

Considered helicoid is the global minimum of the system in the range of fields, $h(\mathrm{n}_2)<h<h(\nu_1)$ (Fig. \ref{PDeta} (b)). In the point n$_2$ it replaces by the first-order phase transition the $+\pi$-skyrmion lattice. Point $\nu_1$ marks the first-order phase transition with homogeneous state. For $h>h(\nu_1)$ such a helicoid can still exist, but as a metastable solution with the positive energy density. In Fig. \ref{PDeta} (a) the region of the helicoid stability is shown by the light violet color.

\subsection{Phase diagram of solutions for $\eta=0.8$}

\begin{figure}
\centering
\includegraphics[width=18cm]{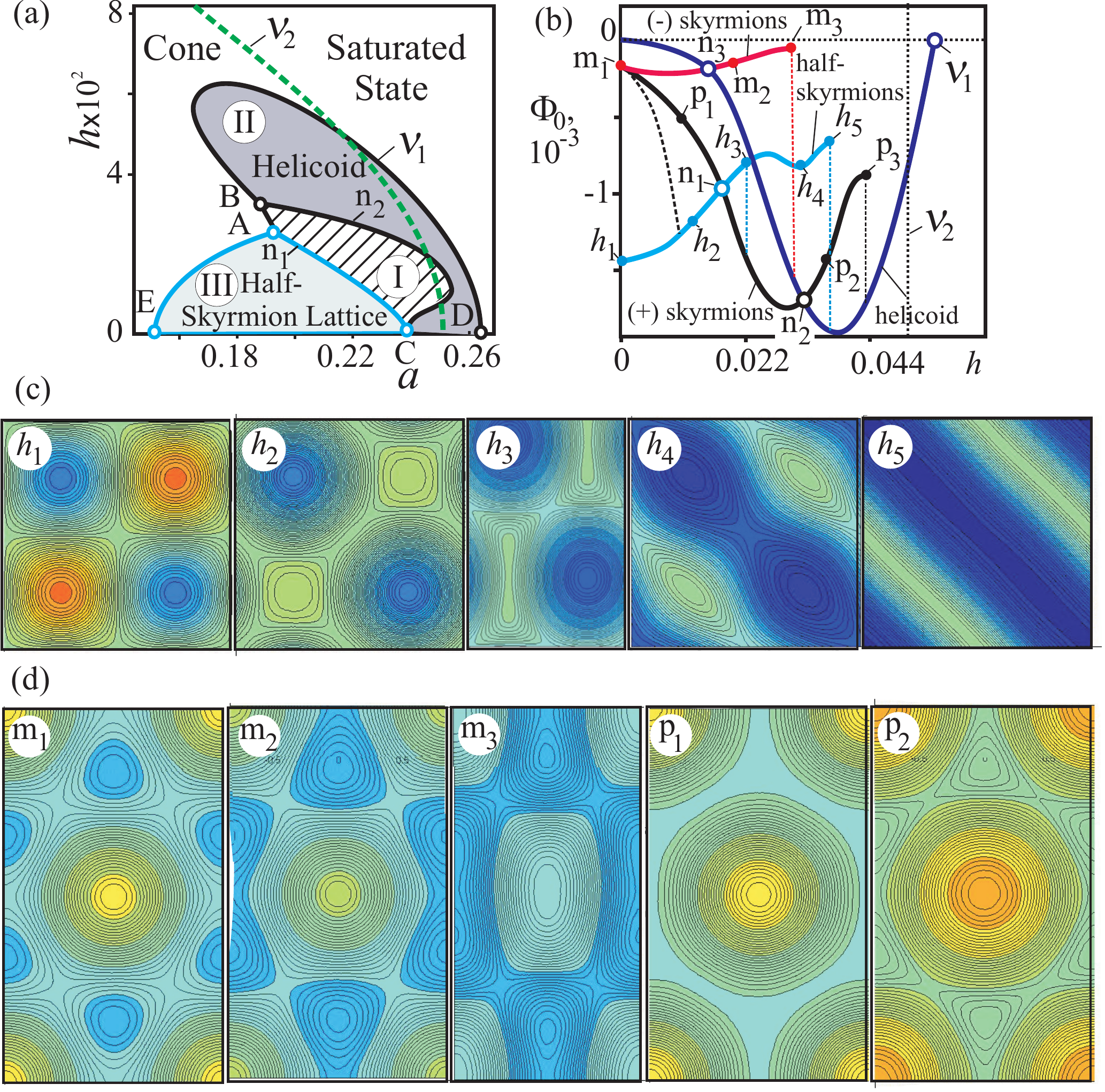}
%
\caption{
\label{PDeta}
(a) Theoretical phase diagram for chiral magnets 
near magnetic ordering according to the modified non-linear sigma-model \cite{Nature06}.
In larger applied fields, i.e. in the A-region, a densely packed
full skyrmion lattice is found in region (I). 
The helicoid transverse to an
applied field is reentrant in region (II).
Region (III) is a half-skyrmion lattice with defects. (b) Dependences of energy densities in all considered modulated phases on the applied magnetic field $h$ ($a=0.23$) calculated with respect to the conical phase. The evolution of skyrmion states is shown in (c) and (d) with the help of contour plots for $m_z$-component of the magnetization. 
}
\end{figure}

The magnetic phase diagram (Fig. \ref{PDeta}) calculated for $\eta =0.8$ includes pockets with square half-skyrmion lattice, hexagonal lattice with
the magnetization in the center of the cells parallel to the applied magnetic field (i.e. $+\pi$ according to terminology introduced in section \ref{skyrmionProperties}), and helicoids with propagation transverse to the field. At low fields, a half-skyrmion staggered lattice is the global minimum of the system. At lines E-A and A-C this lattice undergoes a first-order phase transition into the conical phase and the $+\pi$-skyrmion lattice, correspondingly. At higher field, $+\pi$-skyrmion lattice competes with a helicoidal phase with the line B-C being the line of a first-order phase transition between them.  In contrast, the $-\pi$-skyrmion lattice states expected to form a metastable low-temperature phase in chiral cubic helimagnets (see chapter 4), do not exist near magnetic ordering in this model. Critical points of this phase diagram have the following coordinates: A=(0.209,0.029), B=(0.204,0.036), D=(0.265,0), E=(0.152,0).

The phase diagram shows that both helicoidal kink-like and skyrmionic precursors may exist. 

\section{Conclusions}

In the present chapter, I have investigated the basic phenomenological model for chiral ferromagnets (Eq. (\ref{HTdens})). I 
obtained rigorous solutions for skyrmions  and analytical solutions for one-dimensional helical and conical states in the whole range of the control parameters - the reduced values of temperature, $a$, and magnitude of the applied magnetic field, $h$ (see Eqs. (\ref{coeffa}), (\ref{units1})). I have analysed the transformation of the modulated phases under the influence of the magnetic field and temperature and constructed the phase diagrams of states in Figs. \ref{isolated}, \ref{stabilization}, \ref{PDeta}. 
Here, I highlight the most important results of the present chapter:

(a) By analysing  solutions for localized isolated skyrmions (section \ref{Solutionsconfinement}), it was found that inter-skyrmion coupling being repulsive in a broad temperature range becomes oscillatory near the ordering temperature. This may be explained by "softening" of the magnetization modulus at high-temperatures and strong interplay of angular and longitudinal order parameters. Isolated skyrmions attracting each other may form clusters and confined skyrmion lattices corresponding to minima of skyrmion-skyrmion interaction energy (see Fig. \ref{TwoSk}). Similar effects take place also for helical states \cite{Schaub85,Yamashita87}.

(b) Temperature interval in the phase diagram of Fig. \ref{isolated} may be divided in low- and high-temperature parts: in the main part ($a<a_L=-0.75$, see section \ref{Phenomenonconfinement}) skyrmions are regular chiral modulations with repulsive inter-skyrmion interaction (described in chapter 4). In the high-temperature region ($a>a_L$)  spatial variation of the modulus defines the magnetization processes. 
The confinement temperature $a_L$ is a fundamental parameter of a chiral magnet delineating the border
between two different regimes of chiral modulations. The width of high-temperature interval $\Delta a_2$ is determined by the ratio of isotropic and anisotrpic (DMI) exchange (see Eq. \ref{RationConf}).

(c) Near the ordering temperature skyrmion and helical textures are confined: they can exist only as bound states in the form of clusters or lattices. Trying to push the skyrmions away from the equilibrium (i.e. trying to decrease or increase the period of the skyrmion lattice) leads to the their annihilation (Fig. \ref{expansion}): the farther (closer) the skyrmions from each other the smaller the modulus in the center, and for some critical distance between them, only the homogeneous state is present.

(d) Confined skyrmion and helical textures arise from the disordered state through a rare case of an instability-type nucleation transition.  Decreasing the temperature from paramagnetic region leads to the appearence of skyrmion matter already in the form of lattice. And opposite, the magnetization modulus in skyrmionic lattice gradually decreases to zero with approaching the Curie temperature from the low-temperature part. However, the lattice retains its symmetry up to the critical point.

(e) The properties of confined chiral modulations investigated in this chapter  reveal a noticeable similarity with characteristic peculiarities of cubic helimagnets near the ordering temperatures and known as "precursor states" and
 "A-phase anomalies". This allows to suggest that induced by the softening of the magnetization magnitude the crossover and confinement of chiral modulations is the basic physical mechanism underlying anomalous properties of "precursor states" in chiral magnets.

(f) As the energy differences between different modulated phases in the confinement region are very small,  additional energy contributions result in changes of relative phase stabilities and may cause drastic modification of phase diagrams: cubic anisotropy stabilizes $-\pi$-skyrmions in the particular interval of the magnetic field and half-skyrmions in zero field, whereas in the non-Heisenberg model square half-skyrmions transform into the $+\pi$-skyrmions and eventually into the transversal spirals with increasing magnetic field.

 

\end{document}